\documentclass[]{aiaa-tc}

\usepackage{graphics} % for pdf, bitmapped graphics files
\usepackage{epsfig} % for postscript graphics files
\usepackage{amsmath} % assumes amsmath package installed
\usepackage{amssymb}  % assumes amsmath package installed
\usepackage{cite}
\usepackage{bm}
\usepackage{subfigure}
\usepackage{acronym}
\usepackage{paralist}
\usepackage{float}
\usepackage{color}
\usepackage{epstopdf}
\usepackage{multicol}
\usepackage[]{algorithm2e}
\usepackage{algpseudocode}
\usepackage{tikz}
\usepackage{array}
\usepackage{url}
\usepackage[utf8]{inputenc}
\usepackage[english]{babel}
\usepackage{amsthm}

\usetikzlibrary{automata,positioning}

\newtheorem{Follow field}[Convergence]{Lemma}
\newtheorem{Go-Round Converge}[Convergence]{Lemma}
\newtheorem{Go-Round Safety}[Convergence]{Lemma}
\newtheorem{Change-u Converge}[Convergence]{Lemma}
\newtheorem{Lead-Follow Converge}[Convergence]{Lemma}
\newtheorem{Lead-Follow Safety}[Convergence]{Lemma}
\newtheorem{Mode FT}[Convergence]{Lemma}
\newtheorem{No Zeno}[Convergence]{Lemma}
\newtheorem{Induction N}[Convergence]{Lemma}

\newtheorem{Remark Assum}[Remarks]{Remark}
\newtheorem{Remark CU}[Remarks]{Remark}
\newtheorem{Remark FL}[Remarks]{Remark}
\newtheorem{Remark TD conv}[Remarks]{Remark}
\newtheorem{Remark delta}[Remarks]{Remark}
\newtheorem{Remark Rc Tm}[Remarks]{Remark}
\newtheorem{Remark R14}[Remarks]{Remark}
\newtheorem{Remark Temp Goal}[Remarks]{Remark}

\newtheorem{Assum init}[Assumptions]{Assumption}
\newtheorem{Assum goal}[Assumptions]{Assumption}

\newtheorem{GtG}{Theorem}
\newtheorem{Global Attractive}[GtG]{Theorem}
\newtheorem{Lyapunov}[GtG]{Theorem}
\newtheorem{Input Bound}[GtG]{Theorem}
\newtheorem{Automaton Liveness}[GtG]{Theorem}

\makeatletter
\newcommand{\removelatexerror}{\let\@latex@error\@gobble}
\makeatother

\DeclareMathOperator{\sign}{sign}
\DeclareMathOperator{\F}{\mathrm F}

\DeclareMathOperator{\R}{\mathbb R}

 \author{
  Kunal Garg\thanks{PhD Candidate, Department of Aerospace Engineering, 1320 Beal Avenue, University of Michigan, Ann Arbor, MI 48109. Student Member AIAA.} 
  \ and   Dimitra Panagou 
  \thanks{Assistant Professor, Department of Aerospace Engineering, 1320 Beal Avenue, University of Michigan, Ann Arbor, MI 48109. Member AIAA.} \\
  {\normalsize\itshape
University of Michigan, Ann Arbor, MI, 48109, USA} 
}

\begin{document}

\title{\LARGE \bf Hybrid Planning and Control for Multiple Fixed-Wing Aircraft under Input Constraints}
% \author{Author 1}

\maketitle

\begin{abstract}
This paper presents a novel hybrid control protocol for de-conflicting multiple vehicles with constraints on control inputs. We consider turning rate and linear speed constraints to represent fixed-wing or car-like vehicles. A set of state-feedback controllers along with a state-dependent switching logic are synthesized in a hybrid system to generate collision-free trajectories that converge to the desired destinations of the vehicles. The switching law is designed so that the safety can be guaranteed while no Zeno behavior can occur. A novel temporary goal assignment technique is also designed to guarantee convergence.
We analyze the individual modes for safety and the closed-loop hybrid system for convergence. The theoretical developments are demonstrated via simulation results.    
\end{abstract}

\acrodef{wrt}[w.r.t.]{with respect to}
\acrodef{apf}[APF]{Artificial Potential Fields}
\acrodef{ges}[GES]{Globally Exponentially Stable}
\newcounter{mytempeqncnt}

\section{Introduction}

Distributed control in multi-agent systems has attracted much attention over the last decade, with some of the recent work including but not limited to \cite{parker2009path, mujumdar2011evolving, Ren_Cao_2011,knorn16tcns,wang16neuro, mesbahi10pup}. A fundamental problem of interest is the multi-agent motion planning, i.e., the problem of generating collision-free trajectories for multiple agents so that they converge safely to assigned destinations. More specifically, considering linear speed and turning radius constraints has attracted much attention in recent years \cite{gonccalves2013coordination, nagao2014formation}, motivated in part by conflict resolution for fixed-wing aircraft. Numerous methodologies have been developed, with one of the most popular being vector-field based methods. For instance, \cite{de2015autonomous} utilizes potential-field flow theory, \cite{barry2012safety} presents a barrier function based verification method for reactive controllers, and \cite{roussos20083d} extends the Navigation Function-based approach to 3-D motion. Optimization-based techniques are used in \cite{temizer2010collision, ji2015online}, where the problem of collision avoidance is formulated as a Markov Decision Process. Authors in \cite{kavraki1996probabilistic} present probabilistic maps based method for path planning, see also \cite{karaman2011sampling} for overview of sampling based algorithms for motion planning. The work presented in \cite{shim2005autonomous, stastny2015collision} uses Model Predictive Control (MPC)-based coordination. The survey paper \cite{sujit2014unmanned} presents a comparison among various algorithms for Unmanned Aerial Vehicles (UAVs) path following, such as the carrot-chasing algorithm, vector-field based path following, Pure-pursuit and LOS-based (PLOS) path following, and Non-linear Guidance Law (NLGL). 

Within the plethora of planning and control methodologies, utilizing a hybrid control framework has its own merits. The paper \cite{antsaklis1998hybrid} discusses how hierarchical structures can help with managing complexity, in the sense that they require less detailed models at higher levels (discrete abstraction). In the context of UAV planning and control, several hybrid or switched systems approaches have been developed; for instance, \cite{khamseh2014decentralized} presents a solution that yields convergence to an objective circular path, \cite{casau2013hybrid} presents a hybrid law for autonomous transition flight, and \cite{oikonomopoulos2008hybrid} proposes a safe hybrid control scheme for nonholonomic vehicles flying through an obstacle environment. There has been lot of work on conflict resolution of two aircraft system, see \cite{kosecka1997generation,tomlin2000game,tomlin1998conflict,dallal2017safety}. The authors in \cite{tomlin2000game} present a method of designing hybrid controllers for safety specifications, see also \cite{tomlin1998conflict}. They work with a \textit{relative configuration model} for the case of two aircraft, and define two modes of operation: namely, follow a straight line course, and follow a half-circle. While their approach is limited to two aircraft, it also requires the aircraft to be able to change their heading angles instantaneously, resulting into non-smooth trajectories.

Compared to existing work, the proposed approach and solution in this paper differs in terms of the number of agents in conflict and the technique of resolving conflicts using analytic controllers. Prior work \cite{kosecka1997generation, tomlin2000game} considers the case of two aircraft in conflict resolution, and generates non-smooth trajectories. On the contrary, we consider the case of $N$ agents in conflict, where $N$ can be arbitrarily large, and the proposed protocol yields smooth trajectories for the agents. When it comes to multi-agent coordination, most of the related work assumes that the agents' dynamics are linear; see for instance \cite{olfati2004consensus}. On the contrary, here we consider $N\geq 2$ agents that are modeled under nonlinear, constrained dynamics. The main contributions are as follows: (i) We synthesize a novel hybrid control protocol that generates safe trajectories for agents with input constraints; (ii) We design a novel temporary goal assignment technique to ensure convergence of all the agents; (iii) The distributed nature of the protocol allows it to accommodate arbitrary large number of agents in conflict at the same time, i.e., our protocol is not restricted to pairwise agent deconfliction only.

% We design the guard conditions of the hybrid system in such a way that the safety is assured even in the presence of communication delay. 
%To counteract for the effect of the unknown delay, we design the proposed hybrid control system; this way we obtain sufficient bounds on the maximum possible delay, so that safety is not violated.

The proposed hybrid system consists of 5 control modes (namely, $Go-Round$, $Follow-Leader$, $Change-U$, $Go-towards-Goal$ and $Loiter-at-Goal$) and a switching law determining how an agent switches between these modes, purely on the basis of geometry of the other agents in its neighborhood. Each mode enforces safe maneuvers among two or more agents while satisfying control input constraints. We design the low-level controllers of each mode, as well as the switching logic, i.e., the guards, the resets and the transitions among the controllers, so that safety can be guaranteed at all times and convergence is eventually achieved. Using tools from switched systems theory \cite{zhao2008stability}, we provide a safety and convergence analysis in the presence of the constrained dynamics. In fact, in order to guarantee that every agent reaches its goal location, we adopt a novel temporary goal assignment technique, which eliminates deadlock situations. Whenever there are other agents near an agent's goal location, the agent is allocated a temporary goal location. In addition, we provide a lower bound on the communication radius in terms of the safety distance and the input bounds of the agent, as well as a lower bound on the minimum turning radius in terms of the safety distance. Our approach provides provably correct and safe feedback control solutions in closed-form, which can deconflict multiple agents subject to input constraints. We furthermore design the control laws in each one of the modes so that the resulting position trajectories are smooth. The main feature of the proposed approach is that it provides a way of resolving conflicts involving a large number of agents, while assuring safety, convergence and guaranteeing no Zeno behavior.

% In this paper we consider the problem of generating collision-free trajectories for multiple agents, that are modeled under nonlinear dynamics and input constraints to resemble fixed-wing aircraft or car-like robots. We define a hybrid system consisting of 5 modes (namely, $Go-Round$, $Follow-Leader$, $Change-U$, $Go-towards-Goal$ and $Loiter-at-Goal$), that can be thought as deconflicting maneuvers among arbitrary number of agents (i.e., not merely pairwise deconfliction), and design the low level control laws for each mode, as well as the guards and transitions among them (switching logic). Using tools from switched systems theory \cite{zhao2008stability}, we provide a safety and convergence analysis. As thus one of the major contributions of this paper is that we provide feedback control solutions in closed-form that can deconflict a large number of agents under input constraints. 

The paper is organized as follows: Section \ref{Modeling} includes an overview of the modeling of the system. Section \ref{Switching Modes} presents the modes and corresponding controllers of the hybrid system. In section \ref{Safety} we present the safety analysis of the individual modes, while section \ref{Convergence} includes the convergence analysis. Section \ref{Simulations} evaluates the performance of the proposed method via simulation results. Our conclusions and thoughts on future work are summarized in section \ref{Conclusions}. 

\section{Modeling and Problem Statement}\label{Modeling}
\subsection{System Description}
Consider $N$ agents $i\in\{1,\dots,N\}$, which are assigned to move to goal locations of position coordinates $\bm r_{gi}=\begin{bmatrix} x_{gi}&y_{gi}& z_{gi}\end{bmatrix}^T$ while avoiding collisions. The motion of each agent $i$ is modeled under unicycle kinematics with input constraints to resemble the motion of a fixed-wing UAV, as:
\begin{subequations}\label{unicycle 3D}
	\begin{align}
	\bm{\dot x}_i=\bm f_k(\bm x_i, \bm u_{ik}, q_{ik}) \Rightarrow & \begin{bmatrix}\dot x_i \\ \dot y_i \\ \dot z_i \\ \dot \theta_i\\ \dot \phi_i \end{bmatrix}=\begin{bmatrix}v_{ik} \cos\theta_i \sin\phi_i\\v_{ik}  \sin\theta_i\sin\phi_i \\ v_{ik} \cos\phi_i\\ \omega_{1ik}\\ \omega_{2ik}\end{bmatrix},\\
	\bm u_{ik} = \bm u(q_{ik},\bm x_i, \bm x^i), \quad v_{min} \leq v_{ik} \leq & v_{max}, \quad |\omega_{1ik}| \leq \omega_{max_1},\quad|\omega_{2ik}| \leq \omega_{max_2}
	\end{align}
\end{subequations}
where $\bm x_i=\begin{bmatrix}\bm r_i^T&\theta_i & \phi_i \end{bmatrix}^T \in X_i \subset \mathbb R^5$ is the state vector of agent $i$, comprising the position vector $\bm{r}_i=\begin{bmatrix}x_i&y_i&z_i\end{bmatrix}^T$ and the orientation $(\theta_i, \phi_i)$ of the agent, $\bm u_{ik}=\begin{bmatrix}v_{ik}& \omega_{1ik} & \omega_{2ik}\end{bmatrix}^T \in \mathcal U \subset \mathbb R^3$ is the control input vector comprising the linear speed $v_{ik}$ and the angular speeds $\omega_{i1}, \omega_{i2}$ of agent $i$. $q_{ik} :\mathbb R_+ \rightarrow Q_i$ is the switching signal which is assumed to be piece-wise continuous function where $Q_i = \{q_{i1},q_{i2}, q_{i3}, q_{i4},q_{i5}\}$ is the set of discrete modes and $k\in \{1, 2, 3, 4, 5\}$. The vector field $\bm f_k(\cdot, \cdot,\cdot):\mathbb R^5 \times \mathbb R^3 \times Q_i \rightarrow \mathbb R^5$ is the vector valued function of the agent dynamics in mode $k$. $\bm x^i$ includes the states of agent $i$ as well as those of its neighbors, i.e. $\bm x^i = [\bm x_i^T, \bm x_{i_1}^T, \dots, \bm x_{i_j}^T]^T$ where $i_l \in \mathcal{N}_i$ for $l\in \{1,2,\dots, j\}$. 

% \textbf{JUSTIFICATION OF 2-D PLANAR MOTION}
In this paper, we restrict the motion of the UAVs to 2-D (or planar) motion. One of the main reasons for this constraint is that we are considering the problem of safe trajectory generation of fixed-wing type UAVs flying in low-altitude urban airfield with restrictions on the airspace available for such operations, particularly in terms of altitude restrictions. It is to be noted that with anticipated increase in the number of vehicles in the airspace, it would be desired to have altitude bands designated to different classes of UAVs depending upon their capabilities. Thus, it is desired to design safe trajectories of the aircraft with fixed altitude constraints. Therefore we assume that $\omega_{2ik}(t) = 0$, $\phi_i(t) = \frac{\pi}{2}$ and $z_i(t) = z_{gi}$ for all $t\geq 0$. In rest of the paper, we denote $\omega_{ik} = \omega_{1ik}$. This leads to following system model:
\begin{subequations}\label{unicycle}
	\begin{align}
	\begin{bmatrix}\dot x_i \\ \dot y_i \\ \dot \theta_i \end{bmatrix} & =\begin{bmatrix}v_{ik} \cos\theta_i \\v_{ik}  \sin\theta_i \\ \omega_{ik} \end{bmatrix},\\
	\bm u_{ik}(t) &= \bm u(q_{ik},\bm x_i(t), \bm x^i(t),\label{u-form}\\
	v_{min} \leq v_{ik}(t) & \leq v_{max},\; |\omega_{ik}(t)| \leq \omega_{max} \label{input bound},
	\end{align}
\end{subequations}
with new state vector $\bm x_i = \begin{bmatrix}x_i & y_i & \theta_i \end{bmatrix}^T\in \mathbb R^3$ and control vector $\bm u_i = \begin{bmatrix}v_{ik} & \omega_{ik}\end{bmatrix}^T\in \mathbb R^2$. For the system \eqref{unicycle}, we design controllers $v_{ik}(t)$ and $\omega_{ik}(t)$ so that the solution $\bm x_i(t)$ is well defined in the following sense. We allow $\omega_{ik}(t)$ to be piece-wise differentiable with finite number of discrete jumps, i.e. $\exists \; n<\infty$ and sequence $t_0<t_1<\cdots<t_n<t_{n+1} = \infty$, such that $\omega_{ik}(t)$ is continuously differentiable for $t\in (t_j,t_{j+1}) \quad \forall j \leq n$ with possible jump-discontinuities at the boundary of the intervals. This renders the state $\theta_i(t)$ piece-wise differentiable and continuous for all $t\geq 0$. Furthermore, we design $v_{ik}(t)$ so that its piece-wise differentiable and continuous $\forall t\geq 0$. With this, we get that the trajectory traced by any agent $i$ defined by $(x_i(t),y_i(t))$ is continuously differentiable. From practical point of view, it is important for generated trajectories to be at least continuously differentiable so that the constrained agents such as fixed-wing type aircrafts can follow such trajectories. Our objective is to design the control law $\bm u_i$ for each agent $i$ so that, while maintaining safe distance $d_m$ from other agents, they reach within $r_c$ distance around its goal location $\bm r_{gi}$ where 
\begin{align}
    r_c = \frac{v_{min}+v_{max}} {2\omega_{max}}
\end{align}
and loiter around it. We say that an agent $i$ has reached to its goal location if $\|\bm r_i-\bm r_{gi}\| = r_c$ and it is loitering around the goal location $\bm r_{gi}$. We assume that each agent $i$ has a circular communication/sensing region $\mathcal C_i$ of radius $R_c$ centered at $\bm r_i=\begin{bmatrix}x_i&y_i\end{bmatrix}^T$, denoted as $\mathcal C_i : \{\bm r\in\R^2 \; | \; \|\bm r - \bm r_i\|\leq R_c\}.$ We denote $\mathcal N_{i} = \{ j \; | \; \bm r_j \in \mathcal C_i \}$ the set of agents which are in conflict with agent $i$, or simply, the neighboring agents of the agent $i$. The safe distance $d_m$ is chosen as $d_m = 2\varrho$. We also make the following assumptions for the goals and initial conditions of the agents:
\begin{Assum goal}\label{assum r_g}
$\|\bm r_{gi}-\bm r_{gj}\|>R_c+ 2r_c, \; \forall \; i\neq j$. 
\end{Assum goal}
\begin{Assum init}\label{assum r_0} 
$\|\bm r_i(0) -\bm r_j(0)\| > R_c, \; \forall \; i\neq j$, where $t = 0$ is the initial time.
\end{Assum init}

\begin{Remark Assum}
Assumption \ref{assum r_g} ensures that the agents do not interact with each other once they both are at their goal locations, while Assumption \ref{assum r_0} is required so that agents are conflict-free at $t=0$. 
\end{Remark Assum}

\subsection{Problem Statement}
Formally, the paper deals with generating control input $\bm u_{ik}(t)$ for each agent $i\in \{1, 2, \cdots, N\}$, such that starting from $\bm r_i(0)$, each agent $i$ under the dynamics \eqref{unicycle} reaches its goal location $\bm r_{gi}$ while maintaining safe distance with any other agent $j$, i.e., $\|\bm r_i(t)-\bm r_j(t)\|\geq d_m$ for all $t\geq 0$. Furthermore, the control design should bee such that the resulting trajectories $(x_i(t), y_i(t))$ are continuously differentiable.

\subsection{Parameter Bounds}\label{com del}
From the analysis as per Lemma \ref{GR safety}, we obtain the lower bound on the minimum turning radius as:
\begin{align}\label{min r}
   r_{min} =  \frac{v_{min}}{\omega_{max}} \geq \frac{1}{2}d_m.
\end{align}
From Lemma \ref{FL safe}, we need that $\frac{R_c-d_m}{v_{max}-v_{min}} \geq\frac{\delta_t}{2}$ where $\delta_t>0$ is a small, positive number. From this, we get the lower bound on the minimum communication radius as $R_c\geq\frac{\delta_t}{2}(v_{max}-v_{min}) + d_m$. Define $R_c$ as
\begin{align}\label{Rc min}
    R_c \triangleq  \frac{\delta_t}{2}( v_{max}-v_{min}) + d_m  + \epsilon,
\end{align}
where $\epsilon>0$ is a small, positive number.

\begin{Remark Rc Tm}
It is worth noting that the lower bound on communication radius $R_c$ is independent of the number of agents $N$ and is only a function of system parameters, such as safety distance $d_m$ and input bounds. 
\end{Remark Rc Tm}

\subsection{Notations}
\begin{figure}[!htbp]
    \centering
    \includegraphics[width=0.6\columnwidth,clip]{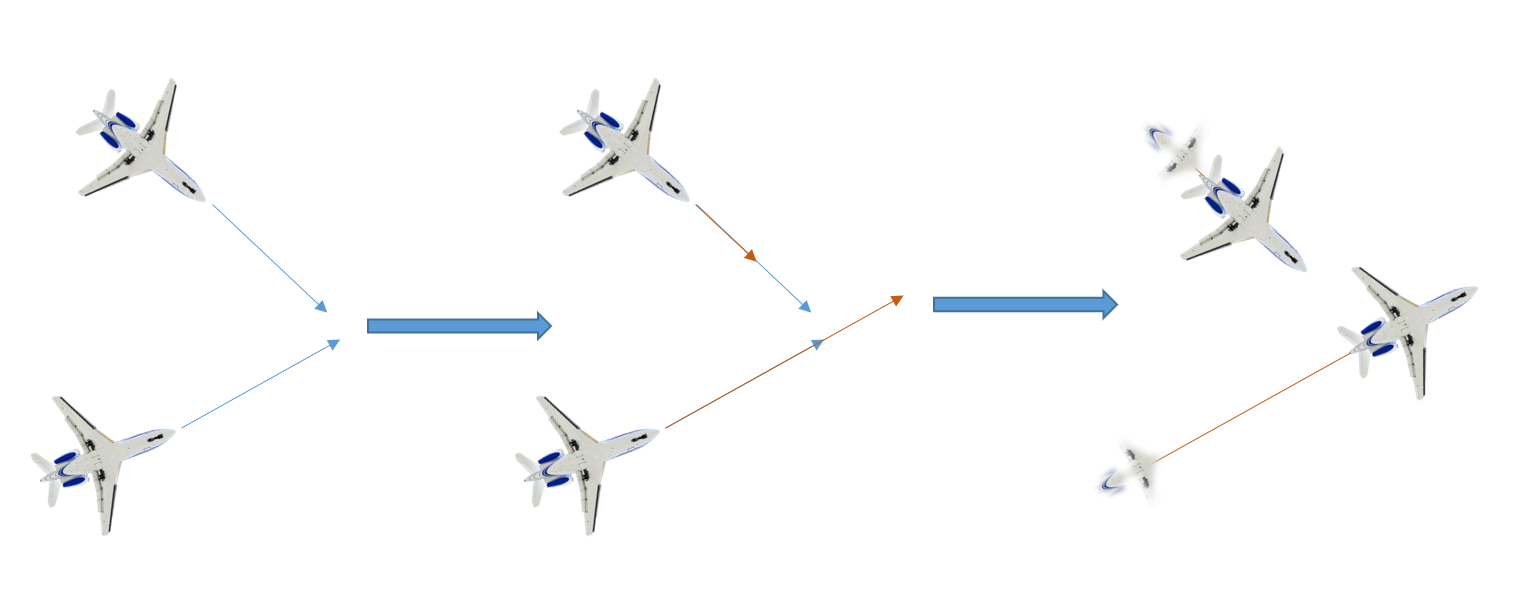}
    \caption{Conflict resolution in Change-u Mode.}\label{fig:CU_mode}
\end{figure}

Throughout the paper, we use $||\bm v||$ for the Euclidean norm of vector $\bm v$, $|v|$ for absolute value if $v$ is a scalar element (e.g., $\theta_i$) and cardinality or number of elements if $v$ is a set (e.g., $\mathcal{N}_i$). We use $\alpha_{ij}$ to denote the difference between $\alpha_i$ and $\alpha_j$, i.e. $\alpha_{ij} = \alpha_i-\alpha_j$. In particular, angular difference between agents $i$ and $j$ denoted as $\theta_{ij}$, is the shortest angle between their orientation vectors, i.e. $\theta_{ij} = \min\{|\theta_i-\theta_j|,2\pi- |\theta_i-\theta_j|\}$. While $\bm r_{gi}$ is the \textit{actual} goal location of agent $i$, we refer to $\bm r_{gi_{temp}}$ as the \textit{assigned} goal location. Parameters $\delta_t, \delta, \Delta$ and $\epsilon$ are small, positive numbers. Lastly, in the Figures \ref{fig:GR_mode}, \ref{fig:CU_mode}, \ref{fig:FL_mode} and \ref{fig:FL_mode2}, the thinner arrows depict the path taken by the respective agent, while the thicker arrows depict the transition of the agents from one mode to another, and their behavior in the new mode. 

\subsection{Design Overview}
We are seeking the synthesis of a hybrid system whose modes accomplish safe trajectory generation and convergence to desired goal locations for multiple agents. The system modes are described in detail in Section \ref{Switching Modes}, and correspond to (combinations of) primitive maneuvers that aircraft-like vehicles can perform, such as moving along a straight line ($M_1$), and moving in a circular path ($M_2$). Table \ref{table:1} gives an overview of the objectives of the various modes, and the situations under which they become active. 

\begin{table}[h!]
\centering
\caption{Overview of the different modes of the hybrid system.}
\begin{tabular}{|m{1cm} | m{5.5cm} | m{5cm}|} 
 \hline
 \textbf{Mode} & \textbf{What} & \textbf{When} \\ 
\hline
 $q_{i1}$ & Go in round-about ($M_2$) & Avoid collision with Agent coming head-on\\ 
 \hline
$q_{i2}$ & Follow Leader: Act as a formation ($M_1$ or $M_2$) & Resolve multiple conflict  \\
\hline
$q_{i3}$ & Change linear speed ($M_1$) & Avoid collision with Agent moving in same direction  \\
\hline
 $q_{i4}$ & Move towards Goal ($M_1$) & No agent in conflict \\
 \hline
$q_{i5}$ & Loiter at Goal ($M_2$) & Once reached at the goal \\ 
 \hline
\end{tabular}
\label{table:1}
\end{table}

Details about the terms used in the hybrid system formulation as well as the control laws are discussed in the following sections. 
\section{Modes of the Hybrid System}\label{Switching Modes}

\subsection{\textbf{Go-Round} ($q_{i1}$)} \label{gr mode}
\begin{figure}[!htbp]
    \centering
    \includegraphics[width=0.5\columnwidth,clip]{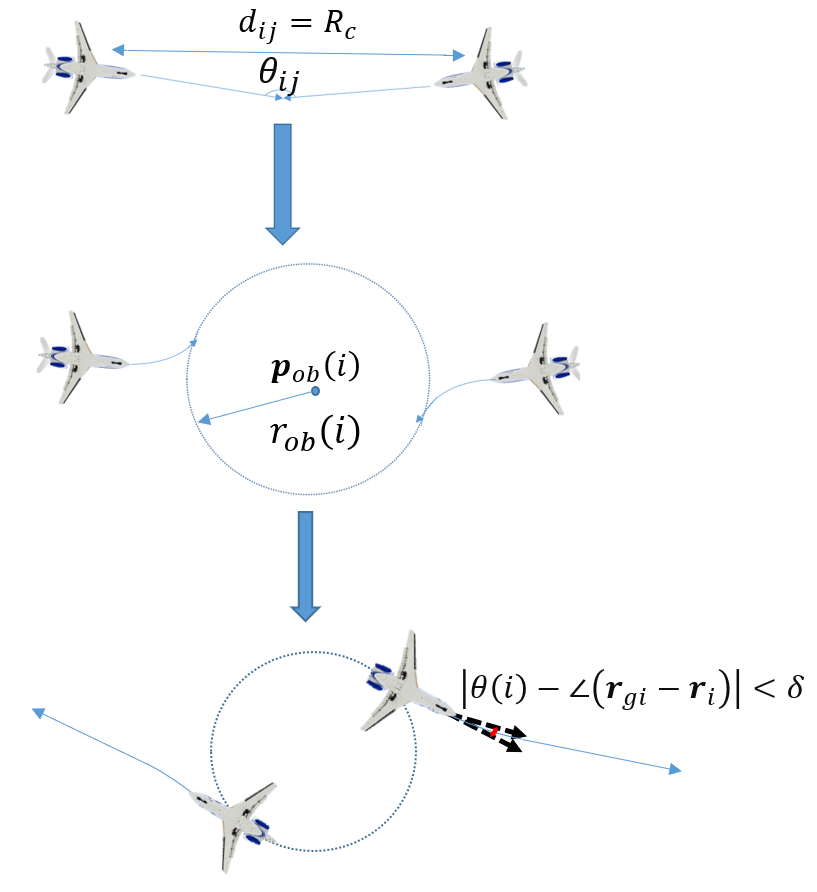}
    \caption{Conflict resolution in Go-Round Mode}\label{fig:GR_mode}
\end{figure}

This mode is used if an agent $i$ is in conflict with another agent $j$ that is not close in terms of orientation, i.e., if their orientation are such that $\theta_{ij} > \theta_c $. In this manner, the agents can maintain safe distance, even with bounded control inputs, by going around a circular path on which their inter-agent distance remains constant. Agent $i$ moves on a circular path $\mathrm{C}_i : \{\bm r \in \mathbb R^2 \; | \; \|\bm r - \bm p_{ob}(i)\| = r_{ob}(i)\} $ whose radius is $r_{ob}(i)$ around the centre $\bm p_{ob}(i) = \begin{bmatrix}p_{obx}(i) & p_{oby}(i)\end{bmatrix}^T$  (see Figure \ref{fig:GR_mode}). 

The center of the orbits $\bm p_{ob}(i) = \bm p_{ob}(j)$ and $r_{ob}(i) = r_{ob}(j)$ are given as: 
\begin{align}
\bm p_{ob}(i) & = \begin{bmatrix}x_i + r_{ob}(i)c\theta_i \\ y_i + r_{ob}(i)s\theta_i\end{bmatrix},\\
r_{ob}(i) &= \frac{v_{ik_i} +v_{jk_j}}{2}t_{min}
\end{align}
where $ t_{min}  = -\frac{\bm r_{ij}^T\dot{\bm r}_{ij}}{|\dot{\bm r}_{ij}|^2}$, $\bm r_{ij} = \bm r_i - \bm r_j$,  $\dot{\bm r}_{ij} = \dot{\bm r}_i - \dot{\bm r}_j$ and $k_i, k_j$ denote the modes of the agent $i$ and agent $j$, respectively, before switching to mode $q_1$. If agent $j$ is already in mode $q_{j1}$, then agent $i$ chooses the same center of the circular orbit but a larger radius, i.e. $\bm p_{ob}(i) = \bm p_{ob}(j)$ and $r_{ob}(i) = r_{ob}(j) + 2d_m$. This allows agent $i$ to maintain the safe distance with agents in the inner circle. 

% \textbf{CHECK IF NEEDED}: If $\mathcal N_i = \emptyset$ and $\bm r_{gi_{temp}} \neq \bm r_{gi}$, then $\bm p_{ob}(i) = \bm r_{gi_{temp}}$ and $r_{ob}(i) = r_c$. (see \ref{sw law} for definition of $\bm r_{gi_{temp}}$).

\begin{figure}[!htbp]
    \centering
    \includegraphics[width=0.5\columnwidth,clip]{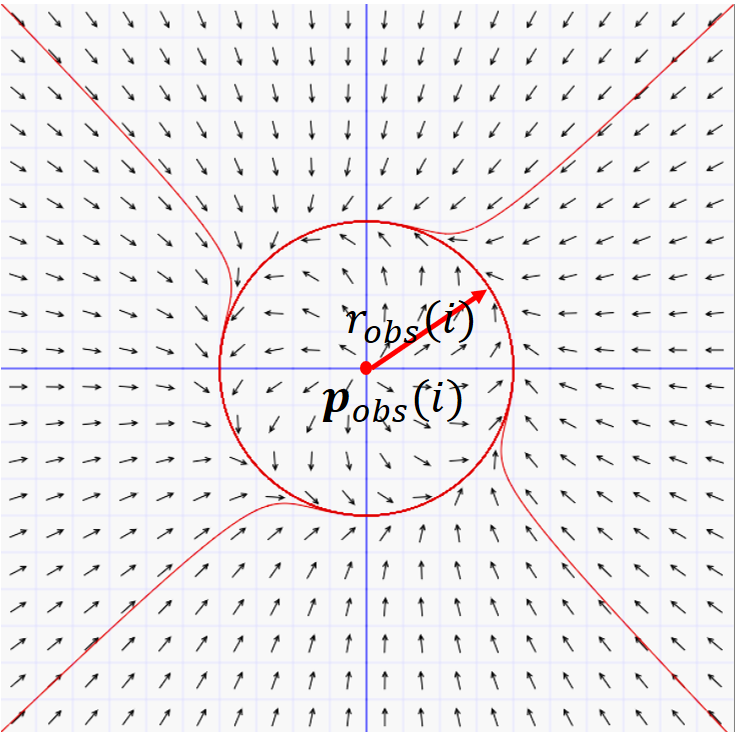}
    \caption{Vector Field for Limit Cycle}\label{fig:Limit}
\end{figure}

The control law $\bm u_{i1}$ under this mode is given as:
\begin{subequations}
\label{go-round}
\begin{align}   
v_{i1} &= k_{v1}(\bm x_{i}(t_s))\sqrt{\F_{i1x}^2(\bm x_i)+\F_{i1y}^2(\bm x_i)}, \\
% \omega_i &= (xu_y-yu_x)/(x^2+y^2),
\omega_{i1} & = -k_{\omega1}(\theta_i - \varphi_{i1}) +\dot{\varphi}_{i1},
\end{align}
\end{subequations}
where $t_s$ is the time instant when agent $i$ switches to mode $q_{i1}$, $k_{v1}$ is given by \eqref{u gain 1}, $\varphi_{i1}  = \arctan\left(\frac{\F_{i1y}(\bm x_i)}{\F_{i1x}(\bm x_i)}\right)$ and the vector field $\textbf{F}_{i1}(\bm x_i) =  \begin{bmatrix} \F_{i1x}(\bm x_i) & \F_{i1y}(\bm x_i) \end{bmatrix}^T$ for the limit-cycle (see Figure \ref{fig:Limit}) is given as:
\begin{align}\label{limit-cycle}
     \textbf{F}_{i1}(\bm x_i)& = \begin{bmatrix} -y + x(r_{ob}^2(i)-x^2-y^2) \\
   x + y(r_{ob}^2(i)-x^2-y^2)\end{bmatrix},
\end{align}
where $x = (x_i-p_{obx}(i))$ and $y = (y_i-p_{oby}(i))$.

The control gain $k_{v1}$ is chosen so that the linear speed is continuous at switching instant, i.e., $v_{i1}(t_s) = v_{ij}(t_s)$ or $k_{v1}(\bm x_{i}(t_s))\sqrt{\F_{i1x}^2(\bm x_i(t_s))+\F_{i1y}^2(\bm x_i(t_s))} = v_{ij}(t_s)$, with $q_{ij}$ being the mode in which agent $i$ was before switching to mode $q_{i2}$. Hence:
\begin{align}\label{u gain 1}
    k_{v1}(\bm x_{i}(t_s)) = \frac{v_{ij}(t_s)}{\sqrt{\F_{i1x}^2(\bm x_i(t_s))+\F_{i1y}^2(\bm x_i(t_s))}}.
\end{align}

\subsection{\textbf{Change-u} ($q_{i3}$)} 
This mode is used to avoid collision between agents coming at a small angular difference. Agent $i$ switches to this mode if $\theta_{ij} \leq \theta_c$ where $j\in \mathcal N_i$ and
\begin{align}\label{critic theta}
    \theta_c & = \arccos\Big(v_r(1-d_r^2)+ \sqrt{(v_r(1-d_r^2))^2-v_r^2-d_r^2(1+v_r^2)}\Big),
\end{align}
where $d_r = \frac{d_m}{R_c}$ and $v_r = \frac{v_{min}}{v_{max}}$.
If the agents are such that one agent is in front of the other agent, in particular, the following geometric condition holds:
\begin{align}\label{cu geom}
    \Big|\theta_i-\arctan{\Big(\frac{y_j-y_i}{x_j-x_i}\Big)}\Big|\leq \arcsin \frac{d_m}{R_c}.
\end{align}
Then, agent $i$ decreases its linear speed and agent $j$ increases its linear speed. If the condition \eqref{cu geom} holds with $i$ replaced by $j$, then agent $i$ increases its speed while agent $j$ decreases its linear speed. If none of these conditions hold, then the agent whose linear speed is smaller at the switching instant decreases it to $v_{min}$, while the other one increases it to $v_{max}$. If the linear speeds of the agents are same, then agent with smaller label value decreases its speed while the other one increases its speed. 
% (see Remark \ref{CU remark}). 

% \begin{figure}[!htbp]
%     \centering
%     \includegraphics[width=0.9\columnwidth,clip]{CU_mode.png}
%     \caption{Conflict resolution in Change-u Mode.}\label{fig:CU_mode}
% \end{figure}

\begin{figure}[!htbp]
    \centering
    \includegraphics[width=0.4\columnwidth,clip]{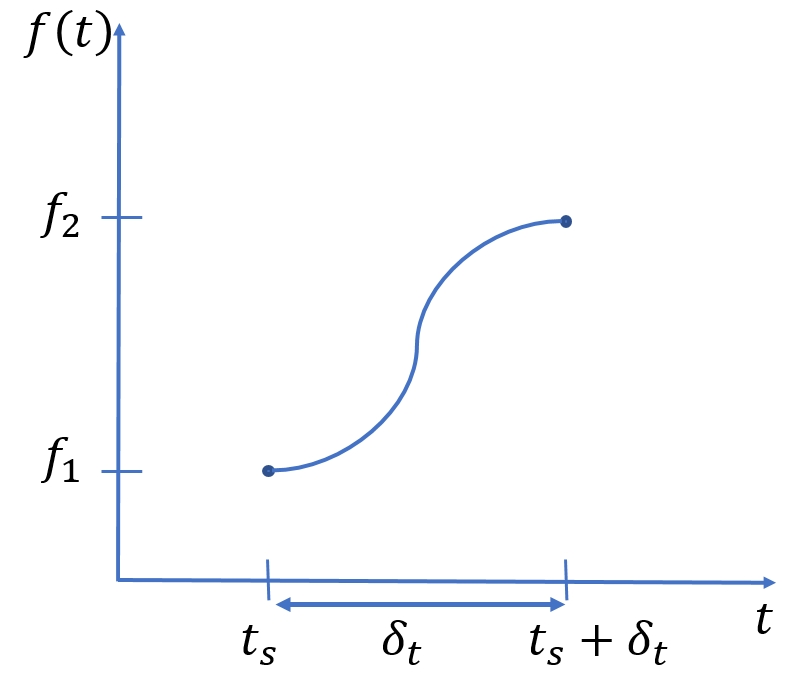}
    \caption{Changing $f(t)$ from $f_1$ to $f_2$ smoothly in time duration $\delta_t$}\label{fig:f var}
\end{figure}

Without loss of generality, assume that agent $i$ decreases and agent $j$ increases its linear speed. The control law $\bm u_{i3}$ is given as:
\begin{subequations}\label{change-u}
    \begin{align}
        v_{i3}(t) &= \left\{
	\begin{array}{ll}
	a_{v3i}s^3+b_{v3i}s^2+c_{v3i}s+d_{v3i}, & \hbox{$s = \frac{t-t_s}{\delta_t}\leq 1$,}\\
	v_{min}, & \hbox{otherwise,}\\
	\end{array}
	\right.\label{c-u-i}\\
\omega_{i3}(t) &= 0.
     \end{align}
\end{subequations}  
Here $a_{v3i},b_{v3i},c_{v3i},d_{v3i}$ are chosen as per \eqref{abcd f} with $f_1 = v_{ik}(t_s)$ where $q_{ik}$ is the mode from which agent switched to mode $q_{i3}$ and $f_2 = v_{min}$ so that $v_{i3}$ smoothly converges to the $v_{max}$ in $\delta_t$ seconds. $t_s$ is the time instant when agent $i$ switches to mode $q_{i3}$ and $\delta_t >0$ is the time duration in which it changes its speed to converge to $v_{min}$. Define $s = \frac{t-t_s}{\delta_t}$ so that $s$ maps the interval $[t_s,t_s+\delta_t]$ to $[0,1]$. $\delta_t$ can be chosen arbitrarily small to have small transient period. We impose the boundary conditions on $f(s) = as^3+bs^2+cs+d$ that $f(0) = f_1 $, $f(1) = f_2$, $\left.\frac{d f}{d s} \right \vert _{s = 0} = \left.\frac{d f}{d s} \right \vert _{s = 1}=  0$, so that the variation of $f(t)$ looks as per Figure \ref{fig:f var}.

With these boundary conditions on $f(s)$, we obtain:
% \begin{subequations}
\begin{align}\label{abcd f}
    \begin{bmatrix} 0 & 0& 0& 1 \\ 0& 0& 1& 0 \\ 1& 1& 1& 1 \\ 3& 2& 1& 0\end{bmatrix}\begin{bmatrix} a \\ b \\ c \\ d\end{bmatrix} = \begin{bmatrix} f_1 \\ 0 \\ f_2\\ 0 \end{bmatrix}
\end{align}
% \end{subequations}
% Similarly, for angular speed we want that $\omega_i$ goes to $0$ smoothly. Hence, we get the coefficients $a_{\omega3i},b_{\omega3i},c_{\omega3i},d_{\omega3i}$ by solving \eqref{abcd f} with $f_1 = \omega_{i4}(t_s)$ and $f_2 = 0$. 

Similarly, the control law of agent $j$ reads:
\begin{subequations}\label{change-u-j}
    \begin{align}        
        v_{j3}(t) & = \left\{
	    \begin{array}{ll}
    	a_{v3j}s^3+b_{v3j}s^2+c_{v3j}s+d_{v3j}, & \hbox{$s = \frac{t-t_s}{\delta_t}\leq 1$,}\\
	    v_{max}, & \hbox{otherwise,}\\
    	\end{array}
	    \right. \label{c-u-j}\\
        % \omega_{j3}(t) & = \left\{
% 	\begin{array}{ll}
% 	a_{\omega3j}s^3+b_{\omega3j}s^2+c_{\omega3j}s+d_{\omega3j}, & \hbox{$s = \frac{t-t_s}{\delta_t}\leq 1$,}\\
% 	0, & \hbox{otherwise,}\\
% 	\end{array}
% 	\right.,
\omega_{j3}(t) &= 0,
    \end{align}
\end{subequations}
where  $a_{v3j},b_{v3j},c_{v3j},d_{v3j}$ are chosen as per \eqref{abcd f} with $f_1 = v_{jl}(t_s)$, where $q_{jl}$ is the mode from which agent $j$ switched to mode $q_{j3}$ and $f_2 = v_{max}$.
% while $a_{\omega3j},b_{\omega3j}, c_{\omega3j},d_{\omega3j}$ are chosen with $f_1 = \omega_{j4}(t_s)$ and $f_2 = 0$. 

% \begin{Remark CU}\label{CU remark}
% Safety of the mode $q_{i3}$ is independent of whether agent $i$ increases or decreases its speed as long as the other agent in the conflict is aware of its control strategy. This heuristic is adopted in order to avoid any uncertainty as to who should increase or decrease their speed.
% \end{Remark CU}

\subsection{\textbf{Follow-Leader} ($q_{i2}$)} 
\begin{figure}[!htbp]
    \centering
      \includegraphics[width=0.7\columnwidth,clip]{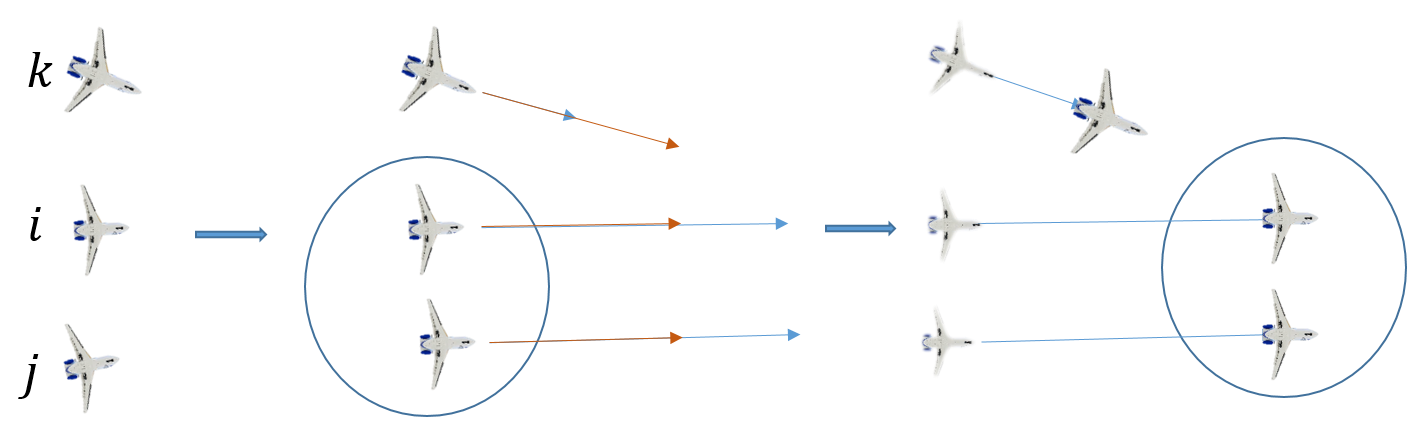}
      \caption{Conflict resolution in Follow-Leader mode via Change-U mode.}\label{fig:FL_mode}
\end{figure}

\begin{figure}[!htbp]
    \centering
      \includegraphics[width=0.7\columnwidth,clip]{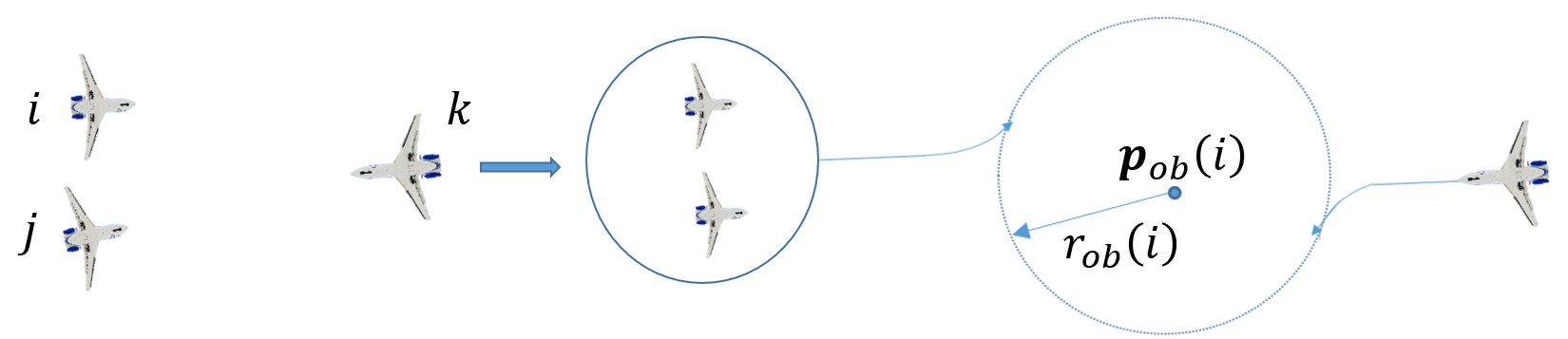}
      \caption{Conflict resolution in Follow-Leader mode via Go-Round mode.}\label{fig:FL_mode2}
\end{figure}

Agent $i$ switches to this mode if it is resolving conflict with agent $j$ while in mode $q_{i3}$, and another agent $k$ comes in conflict with them. Agent $i$ makes a formation with agent $j$ to resolve the conflict with the agent $k$. The formation acts as a single entity and decisions for agents outside the formation are made with respect to the leader $lead(i)$ (see Figure \ref{fig:FL_mode}, \ref{fig:FL_mode2}). For instance, according to the orientation of agent $k$ w.r.t. the leader, the two entities (agent $k$ and the formation) resolve the conflict by switching to either mode $q_{k1}$ or $q_{k3}$. 

While in this mode, agent $i$ follows its leader $lead(i)$, in the sense that it aligns its orientation and linear speed with those of the leader's, while maintaining safe distance from the leader. The agent nearest to the geometric center of the agents is chosen as leader, i.e., $lead(i) = \underset{j \in \textrm{IFW}(i)}{argmin} \|\bm r_j- \bm r_{av}\|$ 
where $\bm r_{av} = \frac{\sum \limits_{j \in \textrm{IFW}(i)} \bm r_j}{|\textrm{IFW}(i)|}$. If there are multiple such $j$, then the agent with the smallest label acts as the leader (see Remark \ref{FL remark}). If agent $i$ is the leader, then it switches to mode $q_{i1}$ or $q_{i3}$ to resolve the conflict with the agent $k$. The control law $\bm u_{i2}$ is given as: 
\begin{subequations}\label{fol-lead}
\begin{align}
v_{i2}(t) & = a_{u1i}s^3+b_{u1i}s^2+c_{u1i}s+d_{u1i}\\
\omega_{i2}(t) &=-k_{\omega2}(\theta_i(t) - \theta_{lead(i)}(t)) + \omega_{lead(i)}(t),
\end{align}
\end{subequations}
where $k_{\omega2}>0$. Here, $a_{u1i},b_{u1i},c_{u1i},d_{u1i}$ are chosen as per \eqref{abcd f} in each interval $[t_j, t_{j+1}]\; j = 0, 1, 2, \dots$, with $f_1 = v_{i2}(t_j)$ and $f_2 = v_{lead(i)}(t_j)$, where $t_0 = t_s$ is the switching instant, and the interval length is $t_{j+1}-t_j = \delta_t$. This ensures that $v_{i2}(t)$ matches with time-varying $v_{lead(i)}(t)$.

\begin{Remark FL}\label{FL remark}
In this case, the choice of leader does not matter as long as all agents know the leader. This heuristic is adopted to avoid any uncertainty as to who should act as the leader.
\end{Remark FL}

\subsection{\textbf{Go-towards-Goal} ($q_{i4}$)} 
Agent $i$ switches to this mode when it is free of any conflict, i.e., $\mathcal{N}_i = \emptyset$. In this mode, agent $i$ moves radially towards its assigned goal location $\bm r_{gi_{temp}}$ under a globally attractive vector field. The control law $\bm u_{i4}$ is as follows:
\begin{subequations}
\label{global-attract}
\begin{align}
    v_{i4} &= k_{v4}(t_s) , \label{u-g}\\
    \omega_{i4} &= -k_{\omega4}(\theta_i - \varphi_{i4}) +\dot{\varphi}_{i4}, \label{w-g}
\end{align}
\end{subequations}
where $k_{v4}(t_s) = v_{ij}(t_s)$, $j$ is the mode in which agent $i$ was before switching to mode $4$, and $t_s$ is the switching instant. $\varphi_{i4}  = \arctan\left(\frac{\F_{i4y}}{\F_{i4x}}\right)$ is the orientation of the vector field $\textbf{F}_{i4}$ which is given by:
\begin{align} \label{global F}
    \textbf{F}_{i4} &= \begin{bmatrix} \frac{-(x_i-x_{gi_{temp}})}{\sqrt{(x_i-x_{gi_{temp}})^2+(y_i-y_{gi_{temp}})^2}} \\
    \frac{-(y_i-y_{gi_{temp}})}{\sqrt{(x_i-x_{gi_{temp}})^2+(y_i-y_{gi_{temp}})^2}}\end{bmatrix}.
\end{align}

\subsection{\textbf{Loiter-At-Goal} ($q_{i5}$)} 
Agent switches to this mode from $q_{i4}$ when it is close enough to its assigned goal location, i.e., if $\|\bm r_i -\bm r_{gi_{temp}}\| = r_c$. In this mode, agent $i$ loiters in a circular orbit centered at $\bm r_{gi_{temp}}$ and of radius $r_c$, under the control law $\bm u_{i5}$ given as:
\begin{subequations}
\label{loiter}
\begin{align}   
v_{i5} &= k_{v5}(\bm x_{i}(t_s))\sqrt{\F_{i1x}^2+\F_{i1y}^2}, \\
% \omega_i &= (xu_y-yu_x)/(x^2+y^2),
\omega_{i5} & = -k_{\omega5}(\theta_i - \varphi_{i1}) +\dot{\varphi}_{i1},
\end{align}
\end{subequations}
where $\F_{i1}$ is given by \eqref{limit-cycle} with $\bm p_{ob}(i) = \bm r_{gi_{temp}}$ and $r_{ob}(i) = r_c$. The gain $k_{v5}(t_s)$ is chosen so that $u_i$ is continuous at the time of switching:
\begin{align}\label{ku5}
 k_{v5} =  \frac{v_{i4}(t_s)}{\sqrt{\F_{i1x}^2(\bm x_i(t_s)+\F_{i1y}^2(\bm x_i(t_s)}}.   
\end{align}

\section{Analysis of Individual Control Laws}\label{Safety}
% In this section, we show the safety and convergence of each mode and that input bounds \eqref{input bound} are satisfied for each agent.
\subsection{Safety and Convergence of mode $q_{i1}$ (\textbf{Go-Round}) and $q_{i5}$ (\textbf{Loiter-At-Goal})}
Since control laws under the modes $q_{i1}$ and $q_{i5}$ are same, we analyze only one of the modes for convergence:
\begin{Go-Round Converge}\label{GR conv}
Under the effect of control law \eqref{go-round}, the closed-loop trajectory of agent $i$ converges to the circular orbit $\mathrm{C}_i$. 
\end{Go-Round Converge}
\begin{proof}
It can be observed that vector field $\textbf{F}_{i1}$ as per \eqref{limit-cycle} has a circular limit-cycle of radius $r_{ob}(i)$ centered at $\bm p_{ob}(i)$ (see Figure \ref{fig:Limit}). Agent $i$ under the control law \eqref{go-round} tracks this vector field asymptotically. This can be verified by choosing the candidate Lyapunov function $V(\theta_i) = (\theta-\varphi_{i1})^2$, whose derivative along the system trajectory under control law \eqref{go-round} reads: $\dot V(\theta_i) = 2(\theta_i-\varphi_{i1})(-k_w(\theta_i-\varphi_{i1}) + \dot \varphi_{i1} -\dot \varphi_{i1}) = -2k_{wi}(\theta_i-\varphi_{i1})^2$. Since $\dot V$ is negative definite, we have that $\theta_i$ tracks $\varphi_{i1}$ asymptotically. Note that the linear speed in control law \eqref{go-round} is non-zero except for $\bm r_i = \bm p_{ob}(i)$. Therefore, agent $i$ follows the vector field \eqref{limit-cycle} and convergse to the circular path around $\bm p_{ob}(i)$. 
\end{proof}

\begin{Go-Round Safety}\label{GR safety}
Under the effect of control law \eqref{go-round}, the agent $i$ maintains safe distance with the agent in conflict if the smallest turning radius is: $r_{min} \triangleq \frac{v_{min}}{\omega_{max}}  \geq \frac{d_m}{2\sin\theta_c}$.  
\end{Go-Round Safety}
\begin{proof}
With the choice of $\bm p_{ob}(i) = \bm p_{ob}(j)$ as per \ref{gr mode}, one has that $v_{i1} = v_{j1}$ in this mode. Therefore, once the agents are on the circle, their inter-agent distance remain constant. The mode is activated only when this constant inter-agent distance is greater than the minimum allowed separation between agents. Agents $i$ and $j$ go to this mode at time instant $t$ only when their angular separation $\theta_{ij} \geq \theta_c$. Now, since the control laws of agent $i$ and $j$ are same under this protocol, their angular difference also remains same (see Figure \ref{fig:GR append}).
\begin{figure}[!htbp]
    \centering
    \includegraphics[width=0.4\columnwidth,clip]{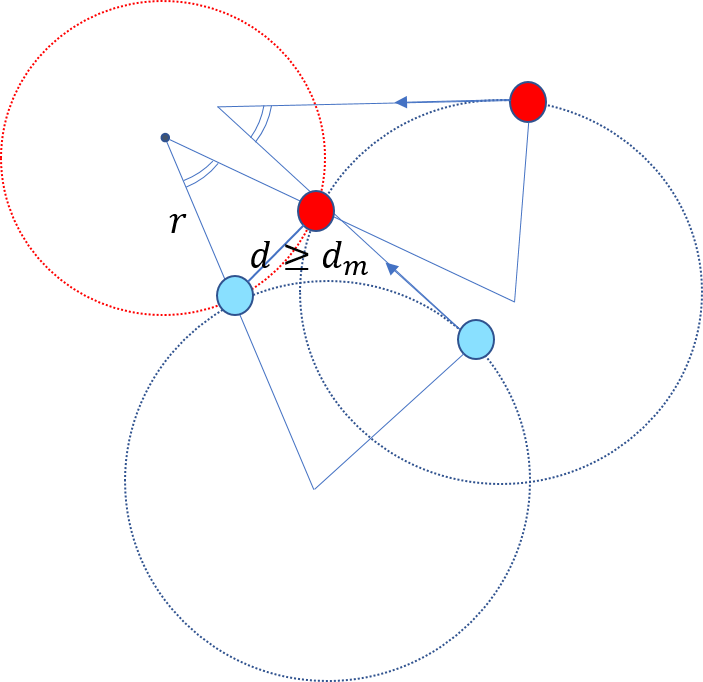}
    \caption{GR mode worst case.}\label{fig:GR append}
\end{figure}

Hence, we have that $d = 2r\sin\theta$, where value of the angle $\theta$ in worst case is $\theta = \theta_c$, while the minimum $r_{min} = \frac{v_{min}}{\omega_{max}}$. Since we impose that $d \geq d_m$, we have that $2r_{min}\sin(\theta_c)\geq d_m  \implies  r_{min} = \frac{v_{min}}{\omega_{max}}  \geq \frac{d_m}{2\sin\theta_c}$. Choosing the minimum radius as \eqref{min r} implies that $q_{i1}$ and $q_{i5}$ are safe.  
\end{proof}

\subsection{Convergence of mode $q_{i2}$ (\textbf{Follow-Leader})}
In this mode, agent $i$ will follow its leader agent $j = lead(i)$ by aligning its linear speed and angular position along those of the \textit{leader} $j$:
\begin{Lead-Follow Converge}
   Under the effect of control law \eqref{fol-lead}, agent $i$ aligns its linear speed $u_i$ and angular position $\theta_i$ along that of leader speed $v_{lead(i)}$ and orientation $\theta_{lead(i)}$, respectively.
\end{Lead-Follow Converge}
\begin{proof}
For the linear speed $u_i$, we observe that under the protocol \eqref{fol-lead}, agent changes its speed to match that of its leader. To simplify the notation, let $j = lead(i)$. Define the error terms as $\theta_{ij} \triangleq \theta_i-\theta_j$. Choose the candidate Lyapunov function as $V(\theta_{ij}) = \frac{1}{2}\theta_{ij}^2$. This function is positive definite and radially unbounded over $\mathbb R$. Taking its time derivative along the closed-loop trajectories, we get $ \dot{V}(\theta_{ij}) = -k_{\omega2}(\theta_{ij})^2$,
which is negative definite over $\mathbb R$. Hence the error term $\theta_{ij}$ asymptotically goes to zero, i.e., agents align its linear speed and orientation with those of its leader. 
\end{proof}
It is also required that the agents maintain safety:
\begin{Lead-Follow Safety}\label{FL safe}
Under the effect of control law \eqref{fol-lead}, agent $i$ maintains a safe distance $d_m$ from its leader $lead(i)$. 
% if $v_{max}-v_{min}\leq 2(R_c-d_m)$. 
\end{Lead-Follow Safety}
\begin{proof}
Without loss of generality, assume that the leader's orientation $\theta_{lead(i)} = 0$ and $v_{lead(i)} = 0$, so that the speed and orientation of agent $i$ are relative to those of the leader's. Assume also that the leader is located at origin, so that we obtain: $\sqrt{(x_i(0)-x_{lead(i)}(0))^2+(y_i(0)-y_{lead(i)}(0))^2} = \sqrt{x_i(0)^2+y_i(0)^2} = R_c$ where $t = 0$ denotes the time instant when the agents detect each other (see Figure \ref{fig:FL_mode safety}). 
\begin{figure}[!htbp]
    \centering
      \includegraphics[width=0.4\columnwidth,clip]{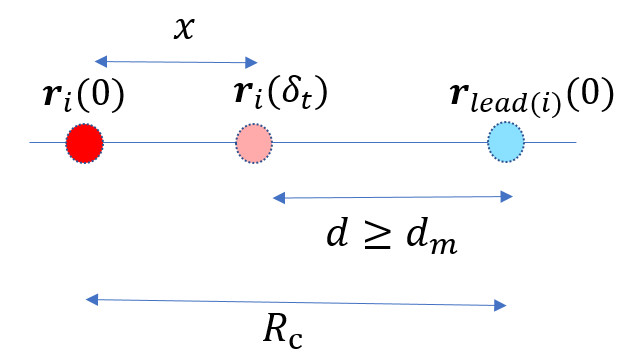}
      \caption{Safety of  Follow-Leader Mode. Dark red point indicate the location of agent $i$ at time instant $t = 0$ when it comes in the communication radius of agent $j = lead(i)$. After switching to mode $q_{i2}$ it moves $x$ distance in the transient mode.}\label{fig:FL_mode safety}
\end{figure}
Once agent $i$ is in the mode $q_{i2}$, as per \eqref{fol-lead} its closed loop dynamics read:
\begin{align*}
    \dot x_i(s) &= (2u_0s^3-3u_0s^2+u_0)\cos0, \quad x_i(0) = x_0, \\
    \dot y_i(s) &= (2u_0s^3-3u_0s^2+u_0)\sin0 = 0, \quad y_i(0) = y_0,
\end{align*}
where $x_0 = -R_c$ and $0\leq s \triangleq \frac{t-t_0}{\delta_t}\leq 1$. By integrating the first equation between $s = 0$ to $1$, we obtain: $ x_i(1) = \frac{u_0}{2}\delta_t + x_0$. From Figure \ref{fig:FL_mode safety}, $x \triangleq |x_i(0)- x_i(1)| = |x_0-(\frac{u_0}{2}\delta_t + x_0)| = \frac{u_0}{2}\delta_t$. For safety, it is required that $d\geq d_m$: 
\begin{align*}
    d = R_c-\frac{u_0}{2}\delta_t \geq d_m  \implies R_c-d_m \geq \frac{u_0}{2}\delta_t  \implies v_{max}-v_{min}\leq \frac{2}{\delta_t}(R_c-d_m).
\end{align*}
Hence with choice of $R_c$ as per \eqref{Rc min}, we obtain that agent $i$ maintains safe distance from its leader while in mode $q_{i2}$.
\end{proof}
\subsection{Safety of mode $q_{i3}$ (\textbf{Change-u})}
This mode is used when the inter-agent angular separation is small, i.e., $\theta_{ij}\leq \theta_c$. In this situation, adjusting the linear speeds of the agents in conflict can maintain the minimum distance:
\begin{Change-u Converge}
    If agents $i$ and $j$ follow the control law \eqref{change-u}, \eqref{change-u-j}, then the inter-agent distance satisfies $d_{ij} \geq d_m$.
\end{Change-u Converge}

\begin{proof}
Under the effect of this control law, agents $i$ and $j$ do not change their orientations. Hence, without loss of generality, we can assume that $\theta_i = 0$. Let $\hat i$ and $\hat j$ denote the unit vectors along the coordinate axes (see Figure \ref{fig:CU_Mode_proof2}).

\begin{figure}[!htbp]
    \centering
        \includegraphics[width=0.5\columnwidth,clip]{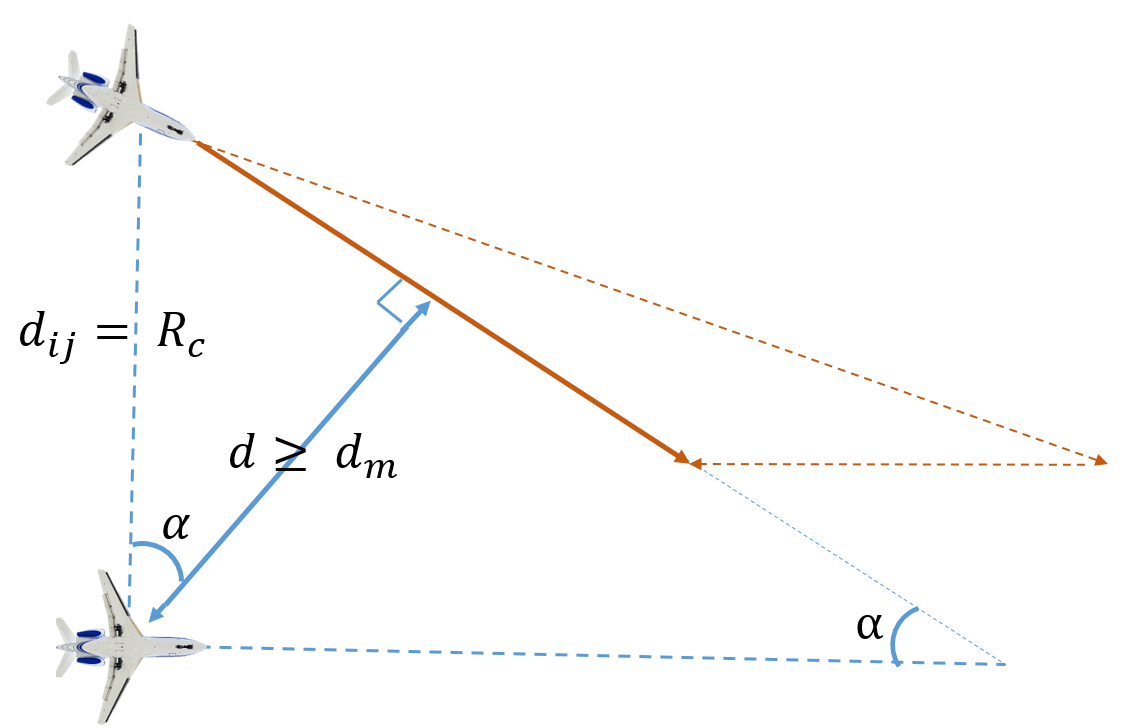}
    \caption{Minimum distance in agent $i's$ frame of reference (relative motion w.r.t. agent $i's$ frame)}\label{fig:CU_Mode_proof2}
\end{figure}
With $\theta_i = 0$, we have:
\begin{align*}
    \bm{\dot{r}}_i = v_i \; \hat i, \quad \bm{\dot{r}}_j = v_j\cos\theta_j \; \hat i + v_j\sin\theta_j \; \hat j.
\end{align*}
If we modify the linear speeds of the agents as per \eqref{c-u-i} and \eqref{c-u-j}, we get:
\begin{align*}
    \bm{\dot{r}}_i = v_{min} \; \hat i, \quad  \bm{\dot{r}}_j = v_{max}\cos\theta_j \; \hat i + v_{max}\sin\theta_j \; \hat j.
\end{align*}

Let the orientation of agent $2$ be $\theta_2 = -\theta$. Let us assume that agent 1 decreases its speed to $v_{min}$ while agents 2 increases its speed to $v_{max}$. From the figure, we need that $d\geq d_m$. We have $d = d_{ij}\cos\alpha$ where $\tan \alpha = \frac{v_{max}\sin\theta}{v_{max}\cos\theta - v_{min}}$. Hence, we need that $\frac{v_{max}\cos\theta - v_{min}}{\sqrt{(v_{max}\sin\theta)^2+(v_{max}\cos\theta - v_{min})^2}}\geq\frac{d_m}{d_{ij}}$. Here $d_{ij} = R_c$ and define $v_r = \frac{v_{min}}{v_{max}}$ and $d_r =\frac{d_m}{R_c}$. From this, we obtain:
\begin{align*}
    & \frac{v_{max}\cos\theta - v_{min}}{\sqrt{(v_{max}\sin\theta)^2+(v_{max}\cos\theta - v_{min})^2}}\geq\frac{d_m}{R_c} 
 \implies \frac{\cos\theta-v_r}{\sqrt{1+v_r^2-2v_r\cos\theta}}\geq d_r\\
    \implies & \cos^2\theta-2\cos\theta v_r(1-d_r^2)+v_r^2-d_r^2(1+v_r^2) \geq 0.
\end{align*}
We thus have $\cos\theta \in [b+\sqrt{b^2-c},1]$, where $b = v_r(1-d_r^2)$, and $c = v_r^2-d_r^2(1+v_r^2)$. Since the agents switch to this mode only if $\theta_{ij}<\theta_c$, from the choice of critical angle $\theta_c$ as per \eqref{critic theta}, we obtain that the agents $i$ and $j$ maintain the safe distance. 
\end{proof}

\subsection{Convergence Analysis of mode $q_{i4}$ (\textbf{Go-towards-Goal})}
\begin{Global Attractive} \label{GA}
\textit{Under the effect of control law \eqref{global-attract}, agent $i$ moves towards its goal location $\bm r_{gi}$.} 
\end{Global Attractive}
\begin{proof}
    Similarly to Lemma \ref{GR conv}, it can be verified that under the control law \eqref{global-attract}, agent $i$ will asymptotically track the vector field \eqref{global F}. Hence, agent $i$ points along the orientation of vector field $\textbf{F}_{i4}$, i.e., $\angle\dot{\bm r}_i = \angle \textbf{F}_{i4}$. From \eqref{u-g}, we have that the magnitude of the velocity vector $|\dot{\bm r}_i| = v_i$. Hence, $\dot{\bm r}_i = v_i\textbf{F}_{i4}$. Choose the candidate Lyapunov function $V(\bm x_i) = \frac{1}{2}(\|\bm r_i - \bm r_{gi}\|^2 + (\theta_i-\varphi_i)^2)$. Taking its time derivative along the system trajectories under the control law \eqref{global-attract} yields: $\dot{V} = (\bm r_i - \bm r_{gi})^T\dot{\bm r}_i + (\theta_i-\varphi_{i4})(\omega_i - \dot \varphi_{i4})$. From \eqref{global F}, the vector field $\textbf{F}_{i4}$ points towards $-(\bm r_i-\bm r_{gi})$ i.e. $\textbf{F}_{i4} = -\frac{(\bm r_i-\bm r_{gi})}{\|\bm r_i-\bm r_{gi}\|}$. Define $\bm r_e = \bm r_i - \bm r_{gi}$ and $\theta_e = \theta_i-\varphi_{i4}$ so that we have:
	\begin{align*}
	    \dot{V} &= (\bm r_i - \bm r_{gi})^T\dot{\bm r}_i + (\theta_i-\varphi_{i4})(\omega_i - \dot \varphi_{i4}) = -\|\bm r_e\| u_i - k_{\omega 4}\theta_e^2\overset{\eqref{u-g}}{=} -k_{v4}\|\bm r_e\|- k_{\omega 4}\theta_e^2.
	\end{align*}
    We have that $\dot{V}$ is a negative definite function over $\mathbb R^3$ and hence, the equilibrium $\bm r_{gi}$ is globally asymptotically stable. Thus, under the effect of control law \eqref{global-attract}, agent $i$ moves towards its goal location. 
    % Now, we show that the agent $i$ reaches to a location $\bm r_i$ such that $\|\bm r_i-\bm r_{gi}\| = r_c$. Note that for any $\bm x\neq 0$, there exists a constant $x_0$, such that $x_0\|\bm x\|\geq \|\bm x\|^2$ for all $\|\bm x\|\leq x_0$. Now, from the above analysis, we know that $\dot V<0$, i.e. the value of the function $V$ decreases for all $t\geq 0$. 
\end{proof}

\subsection{Input bounds in each mode}
To complete the analysis of individual modes, we show that the control input bounds are satisfied in each of the modes: 

\begin{Input Bound}\label{Input Bound}
There exists proper control gains $k_{vj}, k_{\omega j}$ in each mode $q_{j} \; j \in \{1,2,3,4\}$ such that the control input in the respective mode satisfies the constraints \eqref{input bound}, $u_i$ is continuous and piece-wise differentiable, and $\omega_i$ is piece-wise differentiable with finite number of jump discontinuities for $t \in [0,\infty)$. 
\end{Input Bound}
\begin{proof}
See Appendix \ref{App input bound} for input bounds. For the other part, it can be noted that control inputs $v_{ij}, \omega_{ij}$ in each mode $q_{ij}$ are continuously differentiable. Hence, we only need to show that $v_{ij}$ is continuous at the time instant when agent $i$ switches from one mode to another. This can be verified since for each mode $q_{ij}$, the control gain $k_{vj}$ is chosen at the time of switching so that it maintains continuity of $v_{ij}$, i.e., $ v_{i{j_{k-1}}}(t_k^-) = v_{i{j_{k}}}(t_k^+) $ where $t_k$ is the time of switch. For finite number of jumps for $\omega_{ij}$, it is sufficient to show that there there are only finite number of switches, which we prove in Section \ref{Convergence}.   
\end{proof}

In next section, we present the switching logic among the modes of the hybrid system, and prove convergence of the system trajectories to mode $q_5$.
\section{Switching Logic and Convergence to Goal}\label{Convergence}

\subsection{Hybrid System Formulation}\label{sw law}
For each agent $i$, the hybrid system describing the evolution of its state trajectories can be defined using the following \cite{lygeros2004lecture}:

\begin{itemize}
\item The set of discrete states: $Q_i = \{q_{i1}, q_{i2}, q_{i3}, q_{i4},q_{i5}\}$.
\item The set of continuous states: $\bm x = [\bm x_1^T, \bm x_2^T, \dots, \bm x_N^T]^T \in X \subset \mathbb R^{N\times3}$.
\item The vector field: $\bm f_j(\bm x_i, \bm u_{ij}, q_{ij})$ given out of \eqref{unicycle}. 
\item A set of initial states: $X_0 = \{\bm x \; | \; \|\bm r_i -\bm r_j\| > R_c \; \forall j \neq i \} \subset X$.
% \item A domain: $D = \mathbb R^{n\times3}$.
\item A set of edges: $E: Q_i \times Q_i = \{(q_{i0}, q_{i4}), (q_{i1}, q_{i4}), (q_{i4}, q_{i1}), (q_{i2}, q_{i4}), (q_{i3}, q_{i4}), (q_{i4}, q_{i3}), (q_{i3},q_{i2}),\\ (q_{i4},q_{i5}),(q_{i5},q_{i4})\}$.
\item A guard condition $G(\cdot, \cdot): Q_i\times Q_i \to \mathbb R^3$ :
\begin{itemize}
     \item $G(q_{i1},q_{i4}) = \Big\{\bm x_i \in \mathbb R^3 \; | r_{ob}(i) \geq r_{ob}(j)\; \forall j\in \mathcal{N}_i \; \land   \Big(((|\theta_i-\angle(\bm r_{gi} - \bm r_i)|<\delta)\land(\cos(\angle(\bm p_{ob}(i)-\bm r_i)-\angle(\bm r_{gi}-\bm r_i))<0)) \lor (\|\bm r_{gi} - \bm r_i\|< 2r_{ob}(i))\Big) \Big\} $
     \item $G(q_{i4},q_{i1}) = \Big\{\bm x_i \in \mathbb R^3 \; | \exists j \in \mathcal{N}_i \; s.t. \; \Big((\theta_{ij}\geq \theta_c) \lor (AtObstacle(j) = 1) \lor (AtGoal(j) = 1)\Big)\land(A_{41})\Big\} $
     where the condition $A_{41}$ is given as per \eqref{A 41}
     the following inequality:

     \item $G(q_{i3},q_{i2}) = \Big\{\bm x_i \in \mathbb R^3 \; | \; j \in \mathcal N_i, \; \; k\in \mathcal{N}_i \cup \mathcal{N}_j, (k\neq j) \Big\}$
     \item $G(q_{i2},q_{i4}) = \Big\{\bm x_i \in \mathbb R^3 \; | \; \Big(k \in \bigcup_{j\in \textrm{IFW}(i)} \mathcal N_j \Rightarrow k\in \textrm{IFW}(i)\Big) \land \Big((\bm r_{gi_{temp}} - \bm r_i)^T(\bm r_{lead(i)} - \bm r_i)<0\Big) \; \land \Big(\|\bm r_i - \bm r_j\| \geq \|\bm r_k-\bm r_j\| \; \forall k \in \textrm{IFW}(i), \; j = lead(i)\Big) \Big\}$
     \item $G(q_{i3},q_{i4}) = \Big\{\bm x_i \in \mathbb R^3 \; | \; (|\mathcal{N}_i| = 1) \land  \Big(v_{i3}(\bm r_i-\bm r_j)^T\bm \eta_i - v_{j3}(\bm r_i-\bm r_j)^T\bm \eta_j >0 \; j\in \mathcal{N}_i\Big) \Big\}$
     \item $G(q_{i4},q_{i3}) = \Big\{\bm x_i \in \mathbb R^3 \; | \; (|\mathcal{N}_i| = 1) \land \; \Big(\theta_{ij}< \theta_c\; j\in \mathcal{N}_i\Big) \Big\}$
     \item $G(q_{i4},q_{i5}) = \Big\{\bm x_i \in \mathbb R^3 \; | \; \|\bm r_i-\bm r_{gi}\| = r_c \Big\}$
     \item $G(q_{i5},q_{i4}) = \Big\{\bm x_i \in \mathbb R^3 \; | \; (\bm r_{gi_{temp}} = \bm r_{gj})\land (j \in \mathcal N_i )\Big\}$
\end{itemize}
\item A reset map: $R(\cdot, \cdot,\cdot) : Q_i\times Q_i\times X \to \mathbb R^2,\; R(q_{ij}, q_{ij},\bm x^i): \{\bm r_{gi_{temp}} \; | \; (q_{ij}, q_{ij}) \in E\}$ 
\begin{itemize}
    \item $R(q_{i1},q_{i4}) = \Big\{\bm r_{gj_1}\; ;  \; j_1 = \min_{\mathcal N_i} j   \; | \;\Big((\bm p_{ob}(j) = \bm p_{ob}(i))\land (r_{ob}(i)>r_{ob}(j)) \; \forall j \in \mathcal N_i \Big) \land \Big(\|\bm r_{gi}-\bm r_i\|\leq 2r_{ob}(i)\Big) \land (\bm r_{gj} = \bm r_{gj_{temp}})  \Big\} $
    \item $R(q_{i2},q_{i4}) =  \Big\{\bm r_i + r_c \begin{bmatrix}m\sin\theta_i & -m\cos\theta_i\end{bmatrix}^T \; | \; \Big((\bm r_{gi} - \bm r_i)^T(\bm r_{lead(i)} - \bm r_i)\geq 0\Big) \; \land  \Big(\|\bm r_i - \bm r_j\| \geq \|\bm r_k-\bm r_j\| \; \forall k \in \textrm{IFW}(i), \; j = lead(i)\Big), m = \sign(\sin(\angle(\bm r_{gi}-\bm r_i)-\theta_i)) \Big\} $
    \item $R(q_{i5},q_{i4}) = \Big\{\bm r_{gi} +\bm z \; | \; j\in \mathcal N_i \land (\bm r_{gi_{temp}} = \bm r_{gj}) \Big\} $
    \item $R(q_{i4},q_{i4}) = \Big\{\bm r_{gi}\; | \; \mathcal N_i = \emptyset \land (\bm r_{gi_{temp}} \neq \bm r_{gj}) \; j\in \{1, 2, \cdots, N\} \Big\} $
\end{itemize}
\end{itemize}
The inequality $A_{41}$ in $G(q_{i4},q_{i1})$ is given as:
\begin{align}\label{A 41}
     \min_{t, j\in \mathcal N_i}\|\bm r^0_i(t)-\bm r^0_j(t)\|\leq d_m,
\end{align}
where $\bm r_k^0(t) = \bm r_k(t_0) + v_kt\bm \eta_k$ is the position of agent $k$ at time $t$ when it moves with constant speed $v_k = v_k(t_0)$, with $\omega_i(t) = 0$, where $t_0$ is the time when agent $i$ comes in contact with any agent $j$ while in mode $q_{i4}$. 
\subsection{Temporary Goal Assignment}
\begin{figure}[!htbp]
    \centering
    \includegraphics[width=0.35\columnwidth,clip]{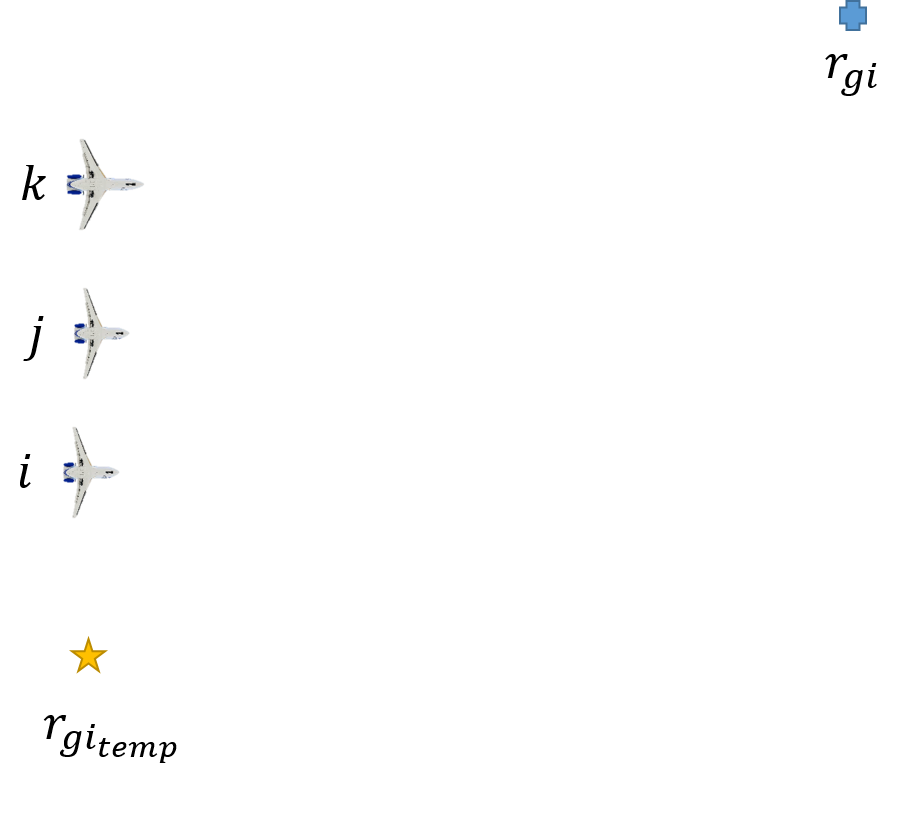}
    \caption{Temporary goal assignment by $R(q_{i2}, q_{i4})$. The "free-side" is the space away from the agents in formation. The temporary goal location is chosen so that it is on the free-side. Here, the blue square is the actual goal location of the agent $i$ while yellow star is the assigned temporary goal location. }\label{fig:temp 24}
\end{figure}

The reset map $R(q_{i1},q_{i4})$ assigns a temporary goal location to agent $i$ when it is in mode $q_{i1}$, and its goal location is very close to its current location. It assigns the goal location of agent $j$ if the agent $j$ is in a circular orbit at the same center as agent $i$ but with a smaller radius, i.e. it is an inner orbit.  This makes agent $i$ go to the goal location of agent $j$. In the reset condition $R(q_{i5}, q_{i4})$, the vector $\bm z$ is a random vector such that $\|\bm z\| = r_c$. This ensures that the temporary goal location and actual location of agent $i$ are different. Th assignment is temporary because as soon as agent $i$ is out of conflicts, i.e. $\mathcal N_i = \emptyset$, the reset condition $R(q_{i4},q_{i4})$ resets the temporary goal location so that $\bm r_{gi} = \bm r_{gi_{temp}}$. Lastly, the reset condition $R(q_{i2}, q_{i4})$ assigns a temporary goal location on the free side (i.e. the space where $(\bm r-\bm r_i)^T(\bm r_j-\bm r_i)<0$ where $j = lead(i)$), so that the agent $i$ can leave the mode $q_{i2}$ (see Figure \ref{fig:temp 24}). 

\subsection{Contents of Communication Package} \label{comm}
Each agent $i$ maintains and communicates certain flags and lists to depict the complete situation in terms of what type of conflict it is in, what control law it is following, and how many and which agents are there around it in conflict:

\begin{itemize}
    \item \textbf{AtGoal} : If the agent $i$ is in mode $q_{i5}$, it sets $AtGoal(i)$ to 1 and keep it 0 otherwise.
    \item \textbf{AtObstacle} : If an agent is in mode $q_{i1}$ or $q_{i5}$, it sets  $AtObstacle(i)$ to 1.  
    % \item \textbf{InFormation}: If agent $i$ is in mode $q_{i2}$, it sets its flag $InFormation(i)$ to 1. 
    \item \textbf{InFormationWith} (IFW): Agent $i$ maintains a list $IFW(i)$ of agents in its formation while in mode $q_{i2}$. 
    % \item \textbf{InConflict}: If an agent $i$ is resolving a conflict with another agent $j$, then both of them will set their flags $InConflict(i)$ and $InConflict(j)$ to 1. 
    \item \textbf{InConflictWith} (ICW): Each agent $i$ maintains a list of the agents $j \in \mathcal N_i$ whom it is in conflict with. 
    \item $\bm p_{ob}$ and $r_{ob}$: Agent $i$ also maintains the position of center $\bm p_{ob}(i)$ and radius $r_{ob}(i)$ of the circular path $\mathrm{C}_i$ while in mode $q_{i1}$ or $q_{i5}$.
\end{itemize}

\begin{figure}[!htbp]
    \centering
    \includegraphics[width=0.45\columnwidth,clip]{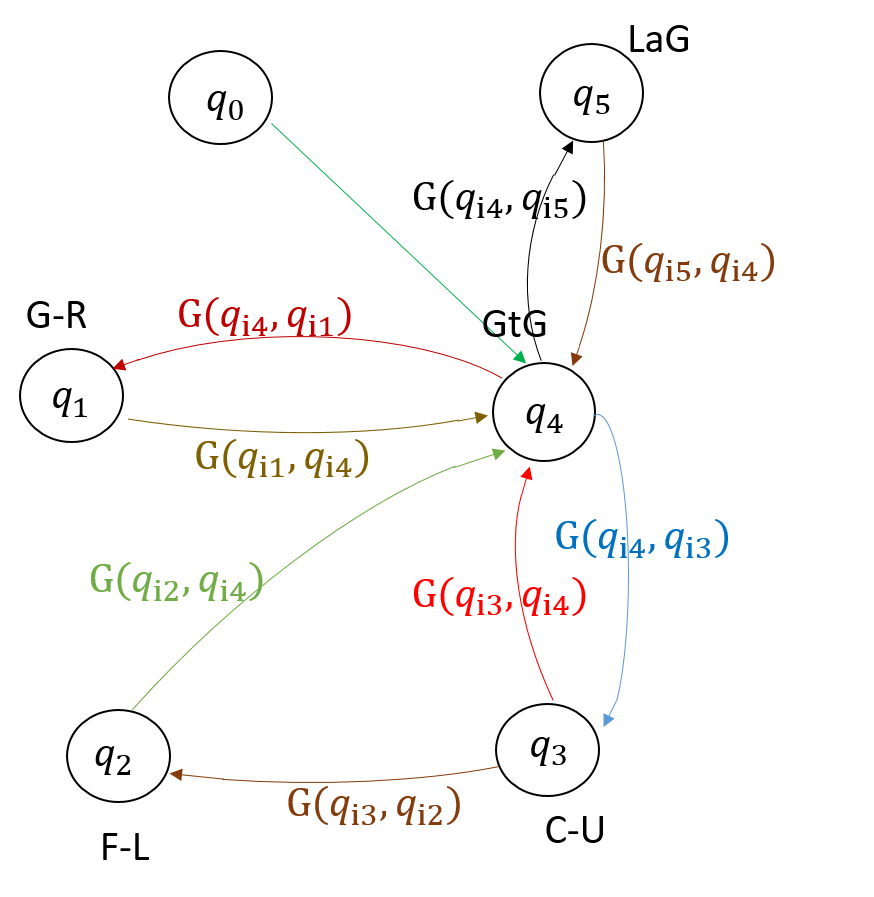}
    \caption{Automaton Representation of the Switching law}\label{fig:autom}
\end{figure}

\subsection{Switching Logic} 
The switching logic of the agents can be represented as an Automaton as per Figure \ref{fig:autom}. Any agent $i$ initiates with mode $q_{i4}$ from any initial condition $(\bm x_i(0), q_{i0})$. Throughout the system evolution, agent $i$ switches between modes as described below:
\begin{itemize}
    \item From Mode $q_{i1}$: It switches to mode $q_{i4}$ if $G(q_{i1},q_{i4})$ is active, i.e., if it does not have any other agents $j\in \mathcal N_i$, with $r_{ob}(i) \geq r_{ob}(j) \; \forall \; j$ at the same centre and if it has its goal in the line of sight, i.e., $|\theta_i-\angle(\bm r_{gi}-\bm r_i)|<\delta$ for some small $\delta>0$. Furthermore, since it switches only if $\cos(\angle(\bm p_{ob}(i)-\bm r_i(t))-\angle(\bm r_{gi}-\bm r_i(t)))<0$, this guarantees that agent does not leave the mode as soon as it enters. Now, if the goal location of the agent lies inside or very close to its circular orbit, i.e., if $\|\bm r_i - \bm r_{gi}\| \leq 2r_{ob}(i)$, then a temporary goal location $\bm r_{gi_{temp}}$ is assigned to the agent $i$ by reset $R(q_{i1},q_{i4})$ to make it switch to mode $q_{i4}$. 
    \item From Mode $q_{i2}$: Agent $i$ switches to mode $q_{i4}$ from $q_{i2}$ when the guard condition $G(q_{i2},q_{i4})$ is satisfied: If there is no agent with which the formation has to resolve the conflict (i.e. all the neighboring agents are in formation), and if the agent is farthest away from the the leader (i.e. $\|\bm r_i-\bm r_j\|\geq \|\bm r_k-\bm r_j\|$), then the agent $i$ leaves the formation by switching to mode $q_{i4}$. If it has its temporary goal location on the free side, i.e., if $(\bm r_{gi_{temp}} - \bm r_i)^T(\bm r_{lead(i)} - \bm r_i)<0 $, it switches to mode $q_{i4}$ directly. Otherwise, it resets its temporary goal location per $R(q_{i2},q_{i4})$ which would fall on the free side and then switch to mode $q_{i4}$.
    \item From Mode $q_{i3}$: Agent $i$ switches to mode $q_{i4}$ when $G(q_{i3},q_{i4})$ is active, i.e. if there is only agent in conflict ($|\mathcal N_i| = 1$) and it starts moving away from the agent in conflict $j$ (i.e. ($v_{i3}(\bm r_i-\bm r_j)^T\bm \eta_i - v_{j3}(\bm r_i-\bm r_j)^T\bm \eta_j>0$). It switches to mode $q_{i2}$ when $G(q_{i3},q_{i2})$ is active, i.e., if another agent $k$ comes in conflict. 
    \item From Mode $q_{i4}$: Agent $i$ switches to mode $q_{i1}$ if $G(q_{i4},q_{i1})$ is active, i.e., if the agent-in-conflict (say, agent $j$) is coming head on i.e., $\theta_{ij}>\theta_c$, or if it is already in mode $q_{j1}$ or $q_{j5}$ so that $AtObstacle(j) = 1$. Otherwise, it switches to mode $q_{i3}$ if $G(q_{i4},q_{i3})$ is active, if the agent-in-conflict is coming towards agent $i$ in such a way that $\theta_{ij}\leq \theta_c$. It switches to mode $q_{i5}$ once it reaches a point such that $\|\bm r_i - \bm r_{gi}\| = r_c$. 
    \item From Mode $q_{i5}$: If $\bm r_{gi} = \bm r_{gj}$ for some $j\neq i$, then once agent $j$ comes in communication radius of agent $i$, i.e., when $j\in \mathcal N_i$, agent $i$ switches to mode $q_{i4}$ and resets its temporary goal location as per $R(q_{i5},q_{i4})$.
\end{itemize}

In the following subsection, we show that every agent eventually converges to the mode $q_{i5}$. Theorem \ref{GA} ensures that all the agents reach their respective goal location once they are in mode $q_5$. Hence, it is sufficient to show that every agent reaches this mode and stays in it indefinitely. In other words: (i) every agent $i$ does not stay in any of the modes $q_{i1}$, $q_{i2}$, $q_{i3}$ or $q_{i4}$ indefinitely, and (ii) every agent does not execute Zeno-behaviour on any of the switching surfaces. 

\subsection{Convergence to mode $q_{i5}$}
\begin{Automaton Liveness}
\textit{
Under the effect of designed hybrid control law as per Section \ref{Convergence}-\ref{sw law}, every agent $i$ reaches its goal location.}
\end{Automaton Liveness}
We first provide the outline of the proof, which follows right after:
\begin{itemize}
    \item First we show that for every agent $i$, the assigned goal location $\bm r_{gi_{temp}}$ is an asymptotically stable equilibrium point using results from \cite{zhao2008stability} (Theorem \ref{Lyap Stability}). 
    \item In order to use the aforementioned result, we show that agent $i$ spends only finite amount of time in any of the modes $q_{ij}$, $j\in \{1,2,3,4\}$ (Lemma \ref{FT mode}).
    \item We then show that there is no Zeno behavior at any of the switching surfaces $S^{i_1i_2}$ (Lemma \ref{Zeno S}).
    \item We complete the proof using induction to show that any number of agents $N$ would eventually reach their respective goal locations (Lemma \ref{Ind N}), without getting stuck at the temporarily assigned goal locations $\bm r_{gi_{temp}}$. In other words, we show that eventually $\bm r_{gi_{temp}} = \bm r_{gi}$ for each agent $i$.
\end{itemize}
\begin{proof}
We say agent $i$ has reached to its goal location when it is in mode $q_{i5}$ with $\bm r_{gi_{temp}} = \bm r_{gi}$. Every agent $i$ would reach its goal location if the goal location is an asymptotically stable equilibrium point. To prove that the goal location is asymptotically stable, we make use of the following result from \cite{zhao2008stability} (Th. 3.9):
\begin{Lyapunov}\label{Lyap Stability}
Suppose that for each $k \in Q_i$, there exists a
positive definite generalized Lyapunov-like function $V_k(\bm x)$ with respect to $\bm f_k$ and the associated trajectory. Then,
\begin{itemize}
    \item[i] The equilibrium of the system \eqref{unicycle} with $\bm u = 0$ is stable if and only if there exist class $G\mathcal K$ functions $\alpha_j$ satisfying
    \begin{align}\label{lyap stab}
        V_j(\bm x(t_{j_{k+1}})) -  V_j(\bm x(t_{j_1})) \leq  \alpha_j(\|\bm r_i(0)-\bm r_{gi}\|), \;  \forall k \geq 1, \;  j = 1, 2, \dots , 4.
    \end{align}
\item[ii] The equilibrium of the system \eqref{unicycle} is asymptotically stable if and
only if \eqref{lyap stab} holds and there exists $j$ such that $V_j(\bm x(t_{j_k}))) \rightarrow 0$ as $k\rightarrow \infty$. 
\end{itemize}
\end{Lyapunov}

Since we are concerned about reaching to the goal, we consider the following common candidate Lyapunov function for any mode of the hybrid system:
\begin{align}\label{cand lyap}
    V_j(\bm r_i) = \frac{1}{2}\|\bm r_i-\bm r_{gi}\|^2 \quad \forall q_{ij}\in Q_i.
\end{align}
From Theorem \ref{GA}, we have that mode $q_{i4}$ satisfies the condition (ii) of Theorem \ref{Lyap Stability}. To prove that any agent $i$ would eventually remain in this mode, we first show that none of agents remains forever in mode $q_{i1}$, $q_{i2}$ or $q_{i3}$. To prove that candidate Lyapunov functions in each mode remain bounded, we show that any agent $i$ spends only finite amount of time in each mode. Lastly, to satisfy the assumption, we show that there is no Zeno behavior on any of the switching surfaces of the hybrid system. 
First we show that agent spends finite amount of time in any mode: 
\begin{Mode FT}\label{FT mode}
\textit{Agent $i$ spends finite amount of time in any mode $q_{ij}$ for $j = \{1,2,3,4\}.$}
\end{Mode FT}
\begin{proof}
Mode $q_{i1}$: Agent $i$ leaves the mode $q_{i1}$ only if $\forall \; j\in \mathcal N_i\; r_{ob}(j)\leq r_{ob}(i)$. If this is not true, every agent $j$ with $r_{ob}(j)>r_{ob}(i)$ would leave the mode. %This leaves the agent $i$ with $r_{ob}(i)\geq r_{ob}(k)$. 
Whenever agent $i$ has its goal in its L.O.S, it will leave the mode. If $R_{11}$ is active, it moves towards a temporary goal to resolve the deadlock. Hence, agent $i$ does not stay in mode $q_{i1}$ indefinitely. 

Mode $q_{i2}$: Agent $i$ remains in the mode $q_{i2}$ only till the formation has an external agent in conflict, i.e., $\exists j \in \mathcal N_i, j \notin IFW(i)$. The formation would switch to either mode $q_{i1}$ or $q_{i3}$ in order to resolve conflict with this agent $j$. As the formation spends only a finite amount of time in any of these modes, it will be free of any conflict in a finite time. Hence, agent $i$ will leave the mode $q_{i2}$ if it is at the maximum distance from the leader. If this is not the case, all the agents farther away from agent $i$ would switch their modes, leaving agent $i$ at the maximum distance from leader. Hence, agent $i$ would eventually leave the mode $q_{i2}$.  

Mode $q_{i3}$: Agent $i$ keeps moving along its current direction, i.e., does not change its orientation while in mode $q_{i3}$. Along the straight line path there always is a point after which the 2 agents in conflict start moving away from each other (i.e., $\dot{d}_{ij} \geq 0$). Beyond that point agent $i$ leaves the mode. The other case is when another agent comes in conflict with the pair. In this case, agent switches to $q_{i2}$.

Mode $q_{i4}$: If any other agent comes in conflict with agent $i$, it would switch to modes $q_{i1}$ or $q_{i3}$. Otherwise, from Theorem \ref{GA}, we have that agent $i$ keeps moving towards its goal location. Hence, in a finite amount of time, it reaches a position such that $\|\bm r_i-\bm r_{gi}\| = r_c$ and switches to mode $q_{i5}$.  
\end{proof}

In summary, agent $i$ cannot stay in any of the modes indefinitely. We now analyze infinite switching at the switching surface $S^{i_1i_2}(i)$, where $i_1$ denotes the initial and $i_2$ denotes the target mode for agent $i$. 
\begin{No Zeno}\label{Zeno S}
\textit{There is no Zeno (chattering) behavior on any of the Switching surfaces for the hybrid system as defined in Section \ref{Convergence}-\ref{sw law}.} 
\end{No Zeno}
\begin{proof}
See Appendix \ref{app: zeno}
\end{proof}

We have so far shown that the system would not stay in any of the modes indefinitely, and would not exhibit any Zeno behavior. To complete the proof, we need to show that the candidate Lyapunov function in each mode remains bounded. Since the chosen candidate function as per \eqref{cand lyap} represents the distance of agent $i$ from its goal location, we need to show that agent travels bounded distance away from its goal location in any of the modes to keep the increment in the candidate Lyapunov function bounded. Define the switching sequences an agent can have, starting from the asymptotically stable mode $q_{i4}$ (see Figure \ref{fig:autom}):
\begin{itemize}
    \item $T_1 = q_{i4}\rightarrow q_{i3}\rightarrow q_{i4}$  
    \item $T_2 = q_{i4}\rightarrow q_{i3}\rightarrow q_{i2}\rightarrow q_{i4}$ 
    \item $T_3 = q_{i4}\rightarrow q_{i3}\rightarrow q_{i2}\rightarrow q_{i1}\rightarrow q_{i4}$ 
    \item $T_4 = q_{i4}\rightarrow q_{i1}\rightarrow q_{i4}$ 
    \item $T_5 = q_{i4} \rightarrow q_{i5} \rightarrow q_{i4}$
\end{itemize}
From Theorem \ref{GA}, we have that the value of the candidate Lyapunov function \eqref{cand lyap} decreases in the mode $q_{i4}$. Agent can have one of the above mentioned switching sequences starting from and ending in this asymptotically stable mode. Hence, we need to show that condition (i) of Theorem \ref{Lyap Stability} is satisfied for all of the above switching sequences.
Assume that an agent $i$ takes $T_l$ switching sequence $n_{il}$ number of times. From the above analysis, we have that agent spends only finite amount of time in any of the modes $q_{ij}$. Let $t_l$ denotes the maximum time that the agent spends in the switching sequence $T_l$ in $n_{il}$ counts. Hence, the worst-case bound of the distance travelled by agent in any sequence $T_l$ is $d_l = n_{il}t_lv_{max}$. This is the maximum distance agent would travel away from its goal location, which is a fixed bounded value depending upon the initial condition $\bm r_i(0)$. Hence, one can choose $\alpha(\|\bm r_i(0)-\bm r_{gi}\|) \in \mathcal K$ such that $d_l\leq \alpha(\|\bm r_i(0)-\bm r_{gi}\|)$. This shows that the hybrid system satisfies both the conditions of the Theorem. 

Now, as per the reset conditions, agent $i$ can be assigned to move towards a temporary goal location. To complete the proof, we need to show that it does not get stuck at a goal location $\bm r_{gi_{temp}} \neq \bm r_{gi}$. 
\begin{Induction N}\label{Ind N}
\textit{Each agent $i$ would eventually reach its own goal location $\bm r_{gi}$. }
\end{Induction N}
\begin{proof}
To prove this, we use induction to show that each agent reaches its own goal location. Case of 1 agent is trivial. For case of 2 agents, we note that 2 agents can resolve their conflict either in mode $q_1$ or $q_3$. In either modes, once agents resolve their conflicts, they would move towards their respective goal location and would not come in conflict with each other. Hence, they would reach their respective goal locations. For case of 3 agents, assume the worst case scenario that all the agents are in mode $q_1$ at the goal location of agent 1. In this case, agent 1 would be assigned to move to the goal location of either agent 2 or 3 depending upon the reset condition $R(q_{i1},q_{i4})$. Lets assume that the temporary goal location of agent 1 is assigned as the goal location of agent 2. From the above analysis, agent would reach to the assigned goal location. Meanwhile, other agents would resolve their conflict and move towards their own goal locations. Once agent 2 reaches its goal, agent 1 would get assigned to move towards its actual goal location. Since agent 2 has already reached to its goal location at this time, we now have a system of 2 agents (agent 1 and agent 3). We have already shown that 2 agents can resolve their conflict and reach their respective goal locations. Hence, all 3 agents would reach their respective goal location. To complete the proof by induction, let us assume that $N$ agents would be able to resolve their conflicts and reach their goal location. For the case of $N+1$ agents, note that there are 2 cases possible:
\begin{itemize}
    \item $(N+1)-th$ agent reaches its goal location after resolving the conflict with all other agents. 
    \item $(N+1)-th$ agents reaches the temporary goal location, which is the goal location of some other agent $j$. 
\end{itemize}
In both cases, one of the agents has reached to the assigned goal location, leaving behind a system of $N$ agents. From the assumption, we have that these $N$ would eventually reach their respective goal location. So, even for the second case above, once agent $j$ reaches its goal location, $(N+1)-th$ agent would reset its goal location first as per $R(q_{N+15},q_{N+14})$ and then as per $R(q_{N+14},q_{N+14})$ to move towards its own goal location. Note that once agent $j$ reaches its goal location, we again have a system of $N$ agents. So, from the assumption of the induction, we have that these $N$ agents would reach their respective goal locations. Hence, we have that all $N+1$ agents reach their goal locations. 

The reset conditions $R(q_{i1},q_{i4})$, $R(q_{i4},q_{i4})$ and $R(q_{i5},q_{i4})$ are carefully designed so that the agents do not keep switching the goals. Since $R(q_{i1},q_{i4})$ only resets the temporary goal location if $\bm r_{gj} = \bm r_{gj_{temp}}$, i.e., the actual and temporary goal locations of the neighbor agent $j$ are same. This avoids the following situation: Assume agent $i$ gets in conflict with some agent $j$ near the location $\bm r_{gi}$ so that its temporary goal location is assigned as $\bm r_{gj}$. We know that agent $j$ would reach its goal location eventually. Now, once agent $j$ reach its goal location, it comes in contact with agent $i$. If $R(q_{i5},q_{i4})$ would assign the actual goal location of the agent directly as the temporary goal location, then $R(q_{j1},q_{j4})$ would be satisfied for agent $j$ and hence, the agent $j$ would be assigned to move to the goal location of the agent $i$. This back-and-forth motion can happen indefinitely for some specific set of initial conditions. On the other hand, the reset condition $R(q_{i5},q_{i4})$ assigns a temporary goal location slightly different from the actual goal location of the agent. This prevents the above situation from occurring as the conditions for the reset $R(q_{j1},q_{j4})$ are not met for the agent $j$. Furthermore, the condition $R(q_{i4},q_{i4})$ makes sure that if and only if the agent is free of any conflict, its temporary goal location is reset as its actual goal location. 
\end{proof} 
This shows that for a system of $N$ agents, where $N$ can be arbitrary, all the agents would be able to resolve their conflicts while maintaining safety (section \ref{Safety}), and would eventually reach their goal locations. 
\end{proof}

% This shows that for a system of $N$ agents, where $N$ can be arbitrary, all the agents would be able to resolve their conflicts while maintaining safety (section \ref{Safety}) and would eventually reach their goal locations. Now we present a simulation scenario to show the performance of the hybrid system.

\section{Simulations}\label{Simulations}
We consider three simulation scenarios involving 10 agents, 20 agents and 2 agents, respectively, with $1.2\leq u_i\leq 1.8$ and $|\omega_i| \leq 0.5$ for all agents. The minimum allowed distance is chosen as $d_m = 0.41$ while the communication radius is chosen as $R_c = 1.64$. With these parameters, conditions of Lemma \ref{GR safety} and \ref{FL safe} are satisfied. 
\subsection{Simulation with 10 agents}
In the first scenario, the initial and goal locations of the agents are given in the Table \ref{table:Sim Case 1}. These are chosen in such a way that agents encounter lots of cross-overs during their motion. Thus many conflicts are created during the system evolution among the agents, showing the efficacy of the designed protocol in handling them.   

\begin{table}[h!]
\centering
\caption{Initial and Goal Location for Case 1.}
\begin{tabular}{|m{0.5cm} | m{2.1cm} | m{2.1cm}| m{0.5cm} | m{2.1cm} | m{2.1cm}|} 
 \hline 
 $i$ & $\bm r_i(0)$  & $\bm r_{gi}$ & $i$ & $\bm r_i(0)$  & $\bm r_{gi}$ \\ 
\hline
 $1$ &(25, 37.5) & (-30, -60) & 6 &(-25, -37.5) &(30,60) \\ 
 \hline
  $2$ &(50, 20) & (-60, -30) & 7 & (-50, -20)& (60, 30)\\
 \hline 
 $3$ &(50, 0) & (-60, 0) & 8 &(-50, 0)& (60, 0)\\ 
 \hline
  $4$ &(50, -20) & (-60, 30)& 9 &(-50, 20)  & (60, -30)\\ 
 \hline
  $5$ &(25, -37.5) & (-30, 60)& 10 & (-25, 37.5) & (30, -60)\\ 
 \hline
\end{tabular}
\label{table:Sim Case 1}
\end{table}

\begin{figure}[!h]
	\centering
	\includegraphics[width=0.75\columnwidth,clip]{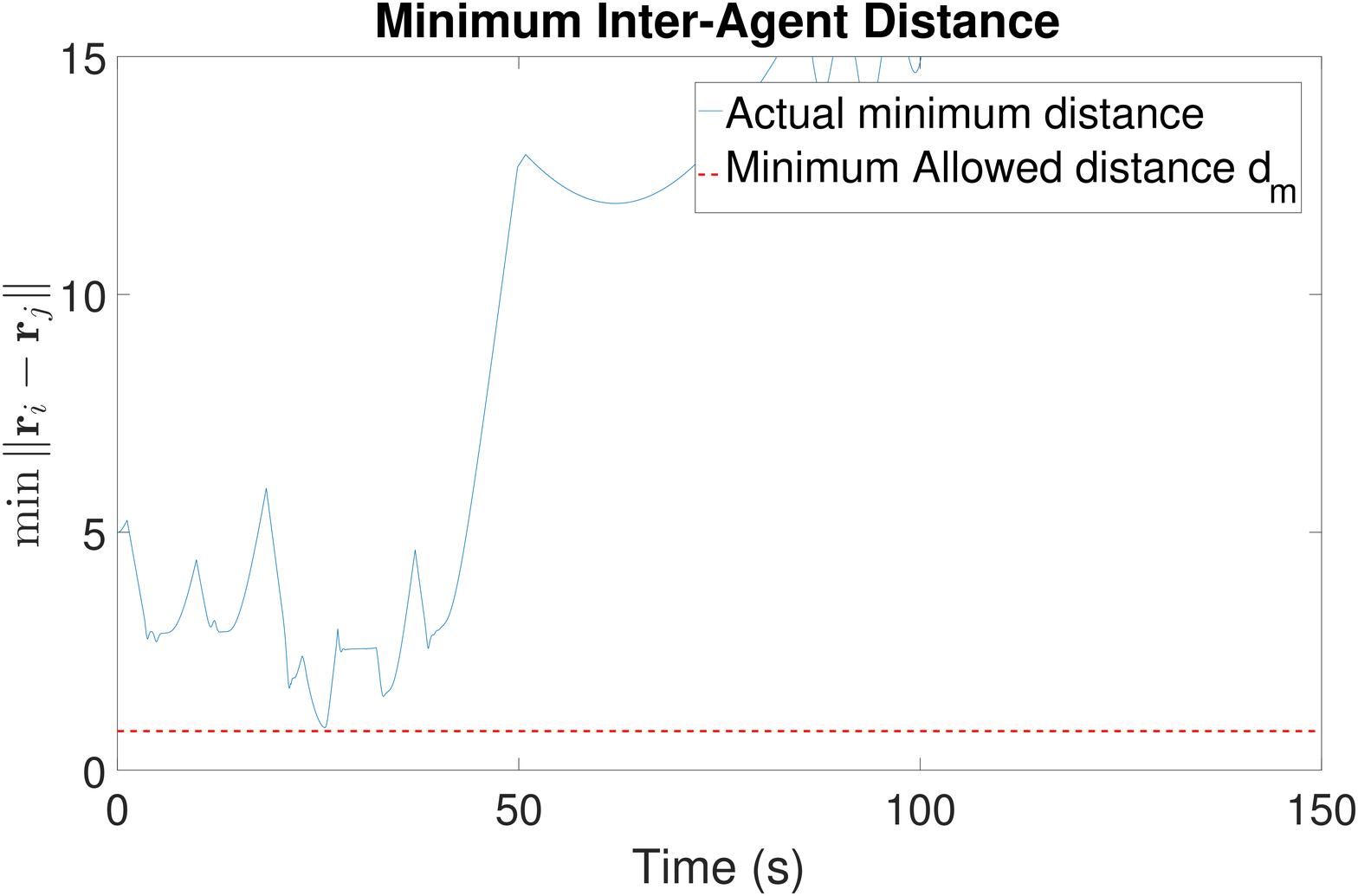}
	\caption{The smallest pairwise distance at each time instant.}
	\label{fig:min dist}
\end{figure}

\begin{figure}[!h]
	\centering
	\includegraphics[width=0.75\columnwidth,clip]{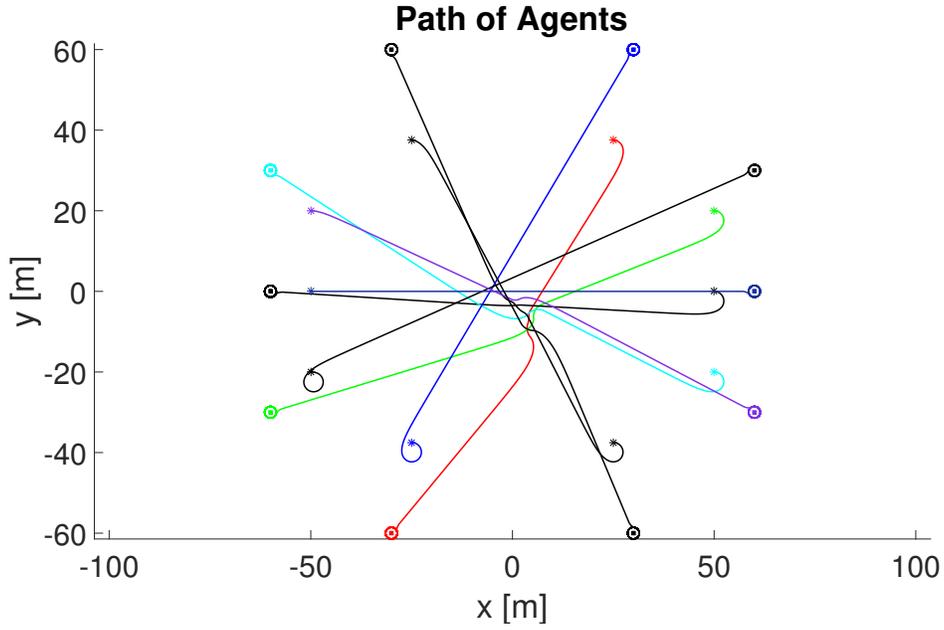}
	\caption{The resulting paths of agents.}
	\label{fig:traj}
\end{figure}

Figure \ref{fig:min dist} shows the minimum pairwise distance between any two agents at each time instant. Clearly, the agents maintain the required minimum distance at all times. Figure \ref{fig:traj} shows the paths of the agents. The paths are smooth and consist of combinations of circular and straight-line segments. The star marks $^*$ correspond to the starting point, while the goal location $\bm r_{gi}$ are marked by square marks \scalebox{.6}{$^\blacksquare$} at the other ends.

\normalsize

\begin{figure}[!h]
	\centering
	\includegraphics[width=0.75\columnwidth,clip]{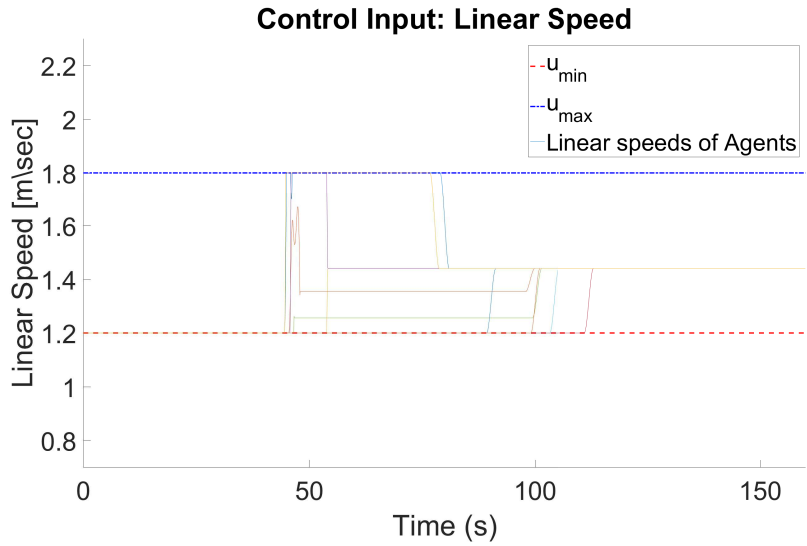}
	\caption{Linear speed $u_i$ of agents with time.}
	\label{fig:lin speed}
\end{figure}

\begin{figure}[!h]
	\centering
	\includegraphics[width=0.75\columnwidth,clip]{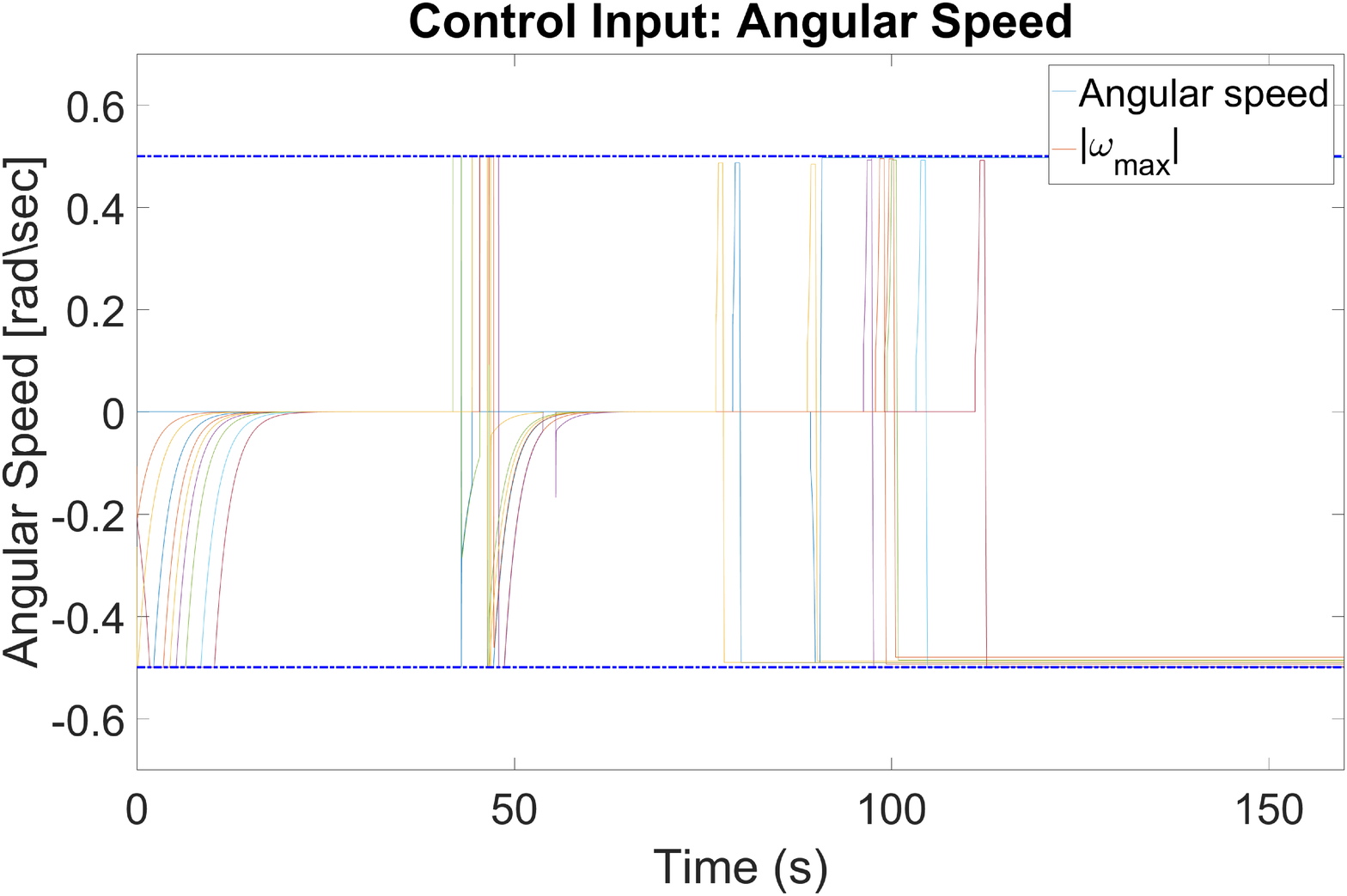}
	\caption{Angular speed $\omega_i$ of agents with time.}
	\label{fig:ang speed}
\end{figure}

Figures \ref{fig:lin speed} and \ref{fig:ang speed} respectively show the evolution of the linear and angular speeds of the agents. It can be seen from the figures that the agents' control inputs are always bounded as per the requirements.

\begin{figure}[!h]
	\centering
	\includegraphics[width=0.75\columnwidth,clip]{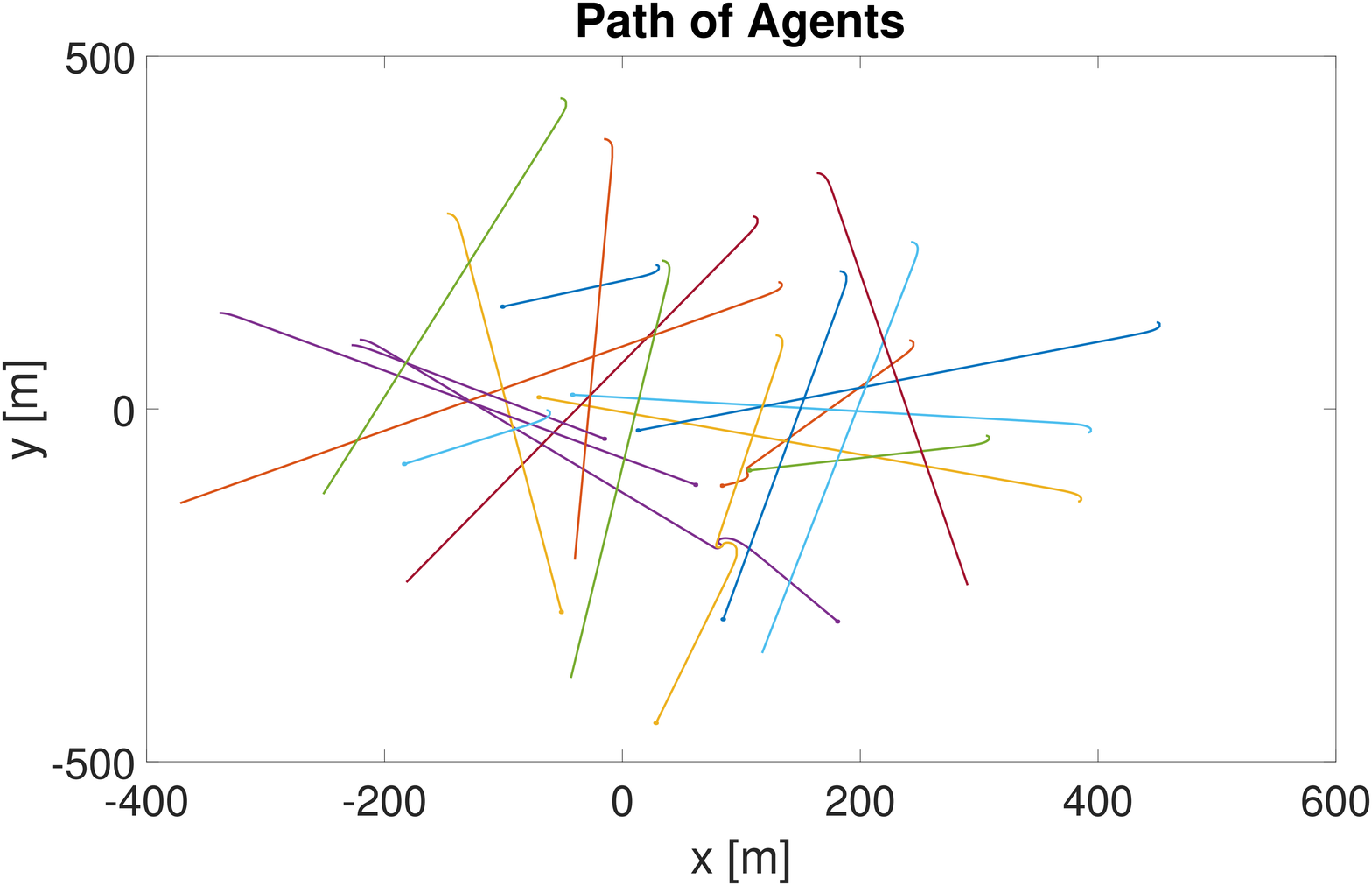}
	\caption{Path of 20 agents.}
	\label{fig:20 path}
\end{figure}

To illustrate that the protocol does not need the symmetry of the initial conditions, goal locations and is not restricted to the case of 10 agents, we performed the simulation with 20 agents with randomly chosen initial and goal locations, satisfying Assumptions 1 and 2. Figure \ref{fig:20 path} shows the path of the 20 agents. All agents reach their desired locations while maintaining safe distances at all times. 

Table \ref{table:time exec dist non-dist actual ref} lists the simulation time $T$ and the normalized time with number of agents $T_N = \frac{T}{N}$ for various number of agents. The simulations were run for 50000 iterations with time step dt = 0.01 sec. As can be seen from the table, our method can be implemented for real-time applications as it is capable of generating trajectories for 500 seconds within 7 seconds of simulation time. 

\begin{table}[h]
\centering
\begin{tabular}{ | m{1cm} |m{2cm} | m{3cm}|} 
 \hline 
 $N$ & $T$ & $T_N$ \\
%  \hline
%   $N_A$ & T/$N_A$(sec) & T/$N_A$(sec)\\ 
\hline
 1 & 6.7951 &  6.7951\\ 
 \hline
 2&  9.2511    & 4.6256    \\ 
 \hline
 3& 10.2750    &  3.4250    \\ 
 \hline
 4& 13.1083   & 3.2771    \\ 
 \hline
 5&   12.7817    &  2.5563   \\ 
 \hline
6& 18.5735    &  3.0956    \\ 
 \hline
 7& 19.8308   & 2.8330    \\ 
 \hline
 8& 18.2672  &  2.2834   \\ 
 \hline
 9&   22.9164   & 2.5463    \\ 
  \hline
 10&  23.4393&  2.3439\\ 
\hline
20 & 44.8112 & 2.2406\\
\hline
\end{tabular}
\caption{Execution Time (in seconds) for various number of agents.}
\label{table:time exec dist non-dist actual ref}
\end{table}

\subsection{Simulation with 3 agents: Illustration of temporary goal assignment}
% \textbf{TO BE DONE}: I will write a para explaining what is happening in this case and put the link of the video. 
% For the second case, the initial and goal locations are chosen such that the goal re-assignment can take place.  First, agents $1$ and $2$ meet at the goal location $\bm r_{g1}$, so that the agent $1$ resets its temporary goal location to that of agent $2$ via $R(q_{11}, q_{14})$. When agents $1$ and $2$ are moving towards the goal location of agent $2$, the agent $3$ meets agent $2$ at the goal location $\bm r_{g3}$ so that agent $3$ also gets assigned its temporary goal location as the goal location of agent $2$ by the reset condition $R(q_{31}, q_{34})$. We have deliberately chosen the speed of agent $1$ more than agent $2$, which makes agent $1$ reaches the $\bm r_{g2}$ before agent $2$. So, the agent $1$ reaches $\bm r_{g2}$ before agent $2$ and stays until the agent $2$ reaches there. Once agent $2$ reaches there, the agent resets its goal location via $R(q_{14}, q_{15})$ and start moving towards the location $\bm r_{g1} + \bm z$. After a small time duration, agent $1$ is free of all conflicts, and hence via $R(q_{14}, q_{14})$, it resets its temporary goal location to the actual goal $\bm r_{g1}$. 
For the second case, the initial and goal locations are chosen such that the goal re-assignment can take place. 
First, agents $1$ and $2$ meet at the goal location $\bm r_{g1}$, so that the agent $1$ resets its temporary goal location to that of agent $2$ via $R(q_{11}, q_{14})$. When agents $1$ and $2$ are moving towards the goal location of agent $2$, the agent $3$ meets agent $1$ at the goal location $\bm r_{g2}$ so that agent $1$ gets its temporary goal location re-assigned as the goal location of agent $3$ by the reset condition $R(q_{11}, q_{14})$. Similarly, agent $2$ also resets its temporary goal location to $\bm r_{g3}$ by $R(q_{31}, q_{34})$. So, all the agents start moving towards $\bm r_{g3}$. 

The speed of agent $1$ and $2$ are deliberately chosen to be greater than that of agent $3$, so that they reach to $\bm r_{g3}$ before agent $3$.  Once agent $3$ reaches there, the agents $1$ and $2$ reset their respective goal location via $R(q_{14}, q_{15})$ and $R(q_{24}, q_{25})$ and start moving towards their own goal locations$\bm r_{g1}$ and $\bm r_{g2}$, respectively. The simulation video for this scenario can be found at the link \url{https://www.dropbox.com/s/rtbn5ircuy0pibg/Sim_3_SciTech.avi?dl=0}.

% After a small time duration, agent $1$ is free of all conflicts, and hence via $R(q_{14}, q_{14})$, it resets its temporary goal location to the actual goal $\bm r_{g1}$. 

% In the Figure \ref{fig:goal re-assign}, the temporary goal assignments of each agents have been plotted against time. When $R(q_{i4}, q_{i5})$ assigns a temporary goal, the goal location of the agent $i$ does not match the actual goal location of any other agent, which is depicted in the Figure by assigning $-1$ value. Similar behavior can be observed for agent $3$ as well: first it is assigned to move to the goal location of agent $2$, then it resets its temporary goal location to $\bm r_{g3} + \bm z$ and eventually, re-assigns the actual goal location.  

\begin{figure}[h!]
	\centering
	\includegraphics[width=0.75\columnwidth,clip]{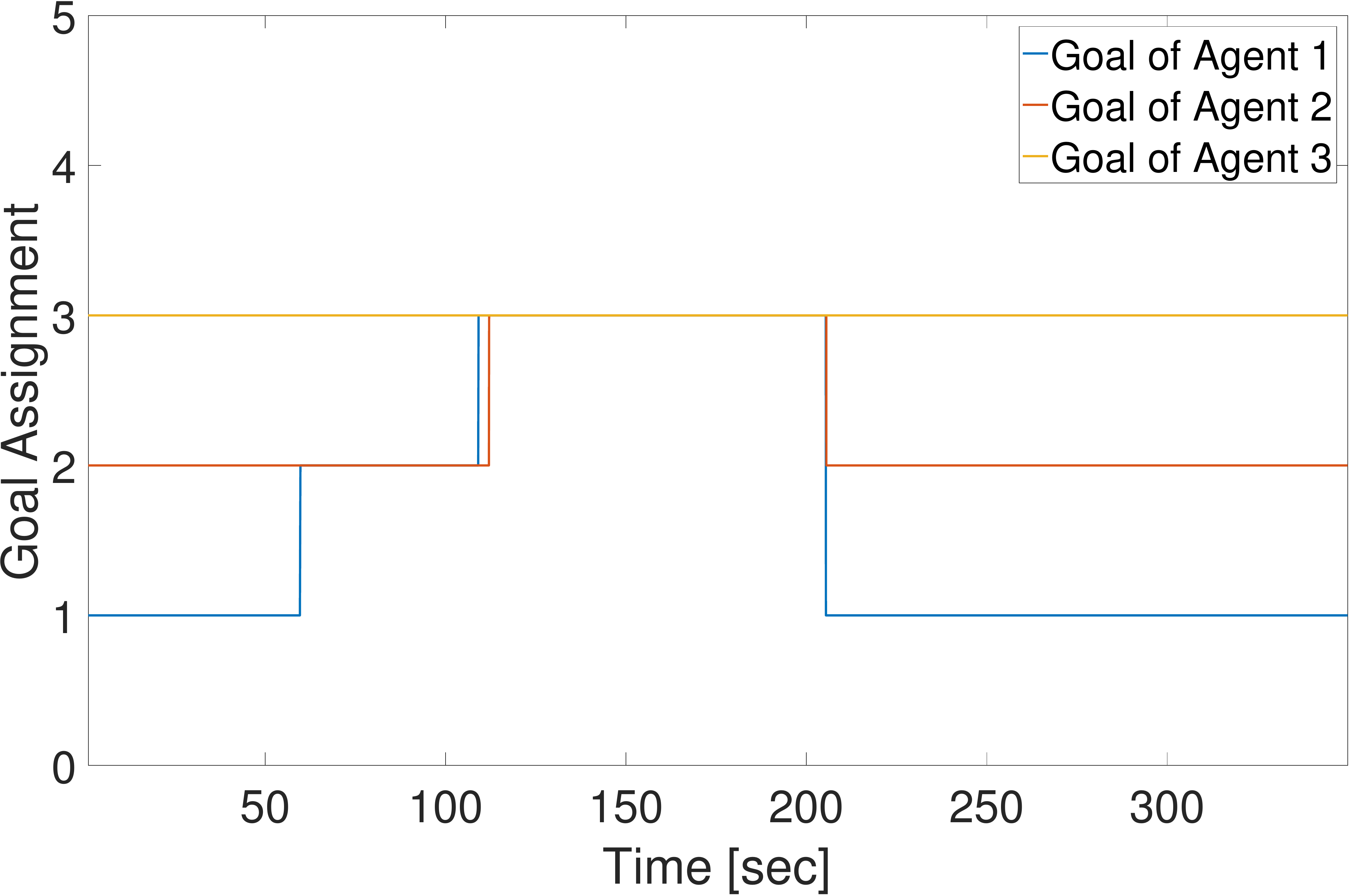}
	\caption{Temporary goal location assignment.}
	\label{fig:goal re-assign}
\end{figure}

\begin{figure}[h!]
	\centering
	\includegraphics[width=0.75\columnwidth,clip]{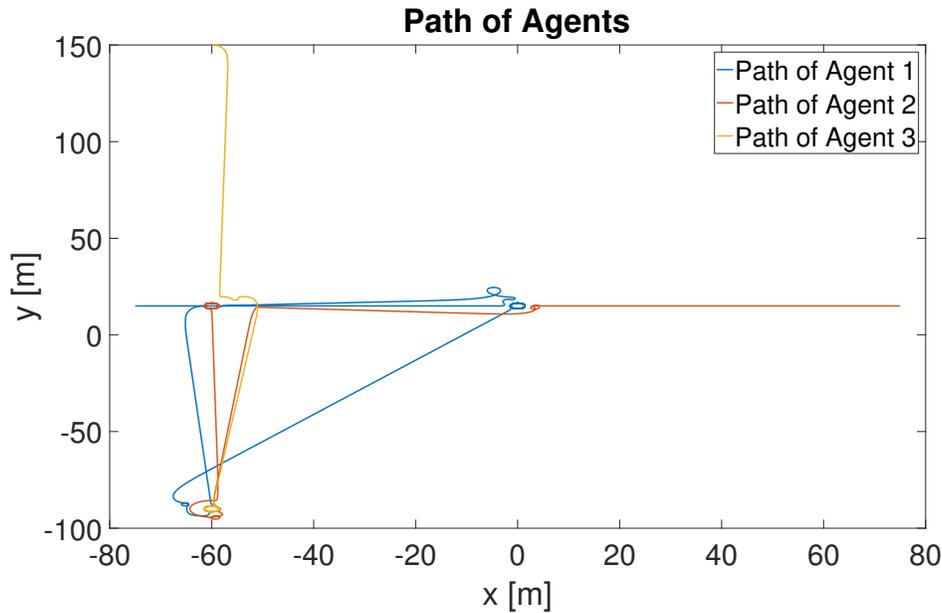}
	\caption{Temporary goal location assignment.}
	\label{fig:goal re path}
\end{figure}

% \subsection{Comparison with MPC based method}
% To demonstrate scalability of our method, we compare the simulation time required for our method with that for Model-predictive control (MPC) based method. We execute the computation using MATLAB R2016a on a desktop with a 16GB DDR3 RAM and an Intel Xeon E3-1245 processor (3.4 GHz). For MPC, we use SQP based distributed nonlinear-MPC with a fixed horizon of $N_p = 5$. 

\section{Conclusions and Future Work}\label{Conclusions}
We presented a safe multi-agent coordination protocol for agents with input constraints, with application to the deconfliction of fixed-wing aircraft. The hybrid protocol is provably safe and scalable with number of agents.  In future, we would like to study the case of input- and state- constrained agents operating in obstacle environments under state and output uncertainties. Our ongoing research focuses on designing hybrid protocols for fixed-wing aircraft in the presence of wind disturbances and sensor noises. In this paper, we assumed that perfect state information is available to the agents, and that there are no communication lags or losses. It would be  interesting to see how the design of the hybrid system changes if the communication between the agents is establishes at discrete time events, i.e., the information is transmitted/received at discrete time instances. 

We demonstrated a way of generating various modes and a hand-synthesized switching law so that the safety and convergence can be guaranteed for a particular class of constrained dynamical system. We are investigating the systematic way of generating such hybrid systems for more general class of nonlinear, constrained systems.  A broader problem that we would like to address in future is, given a dynamical system, its constraints, the information available to the system and the capabilities of the system, how to synthesize the modes and the switching law for switching among the modes so that the given specifications or objectives can be met. 

\section*{Acknowledgments}
The authors would like to acknowledge the support of the NASA Grant NNX16AH81A and the Air Force Office of Scientific Research under award number FA9550-17-1-0284.

\bibliographystyle{aiaa}
\bibliography{myreferences}

\appendix
\section*{Appendix}

\section{Proof of Theorem \ref{Input Bound}}\label{App input bound}
\begin{proof}
Let $\Delta$ be a small, positive number satisfying $0<\Delta <\omega_{max}\frac{v_{min}}{v_{min}+v_{max}}$. We show that for each mode $q_{ij}$, there exist positive gains $k_{vj}$ (wherever applicable) and $k_{\omega j}$ such that the control input gains are satisfied:

\textbf{Mode 1 }: From the symmetry of vector field \eqref{limit-cycle}, we have that that maximum value of both $\|\mathbf F_{i1}\|$ and $\dot\varphi_{i1}$ occurs at the maximum distance from $\bm p_{ob}(i)$. From the choice of $k_{v1}$ as per \eqref{u gain 1}, we can bound $v_{i1}$ as $ \frac{r_{ob}(i)}{\|\mathbf F_{i1}\|}v_{i4}\leq v_{i1}\leq v_{i4}$, which implies $v_{i1}\leq v_{max}$ since $v_{i4}$ already satisfies the control bounds. It is sufficient to choose $r_{ob}(i) \geq v_{min}\|\mathbf F_{i1}(\bm x)\|$, where $\|\bm x-\bm p_{ob}\| = R_c$ so that $v_{i1} \geq v_{min}$. With this choice of $r_{ob}(i)$, we have that $v_{min} \leq v_{i1}$. Also, we can bound the angular speed as $\omega_i \leq 2k_{\omega1}\pi + |\dot\varphi_{i1}(\bm x)|$. Since the maximum is achieved at $\bm x$ such that $\|\bm x-\bm p_{ob}(i)\| = R_c$, we can find $k_{\omega1}$ such that
\begin{align}
    k_{\omega1} \leq \frac{1}{2\pi}(\omega_{max} - |\dot\varphi_{i1}(\bm x)|-\Delta),
\end{align}
so that $|\omega_{i1}|\leq\omega_{max}-\Delta$.

\textbf{Mode 3 }: In this mode, we have $\omega_{i3} = 0$. Also, agent $i$ changes its linear speed to either $v_{min}$ or $v_{max}$ monotonically. Hence, the input constraints are satisfied in this mode.

\textbf{Mode 4 }: From \eqref{u-g}, we have that the linear speed in this mode is same as that in the previous mode. Since the linear speed in the other 3 modes satisfies the constraint \eqref{input bound}, we have that $v_{i4}$ also satisfies its constraints. For the angular speed, we note that $\dot \varphi_{i4} = v_{i4}(\frac{x_i-x_{gi}}{\|\bm r_i-\bm r_{gi}\|^2}\sin\theta_i-\frac{y_i-y_{gi}}{\|\bm r_i-\bm r_{gi}\|^2}\cos\theta_i)$. Hence, we get 
\begin{align*}
\dot\varphi_{i4} \leq \frac{v_{max}}{r_c} = \omega_{max}\frac{v_{max}}{v_{min}+v_{max}}    
\end{align*}
Let $\omega_{\varphi} = \omega_{max}\frac{v_{max}}{v_{min}+v_{max}}$ so that we get $\omega_i = -k_{\omega4}(\theta_i-\varphi_{i4}) +\dot \varphi_{i4} \leq 2k_{\omega4}\pi + \omega_{\varphi}$. Choose $k_{\omega4}$ as
\begin{align}\label{w4 gain}
    k_{\omega4}& \leq\frac{1}{2\pi}(\omega_{max}- \omega_{\varphi}-\Delta)
\end{align}
so that $|\omega_{i4}|\leq \omega_{max}-\Delta$.
   
\textbf{Mode 5} : While in this mode, agent moves with speed $v_{i5} = v_{i4}(t_s)$ as it chooses $k_{v5}$ as per \eqref{ku5}. 
In this mode, $\omega_{i5} \leq 2\pi k_{\omega5} + |\dot \varphi_{i5}|$, where 
\begin{align*}
    \dot \varphi_{i5} & \leq (\cos\varphi_{i5}\F_{ix5} + \sin\varphi_{i1}\F_{iy5})\frac{v_{i5}}{\F_{ix5}^2+\F_{iy5}^2} \leq \sqrt{F_{ix5}^2+\F_{iy5}^2}\frac{k_{v5}\sqrt{F_{ix5}^2+\F_{iy5}^2}}{\F_{ix5}^2+\F_{iy5}^2} = k_{v5} \overset{\eqref{ku5}}{\leq} \frac{1}{r_c}v_{i4}(t_s).
\end{align*}
Since $r_c = \frac{v_{min}+v_{max}}{\omega_{max}}$, one can choose:
\begin{align}\label{w5 bound}
    k_{\omega5} \leq \frac{1}{2\pi}\Big(\omega_{max}(1-\frac{v_{min}}{v_{min}+v_{max}})-\Delta\Big),
\end{align}
so that $|\omega_{i5}|\leq\omega_{max}-\Delta$. 

\textbf{Mode 2} : From \eqref{fol-lead}, it can be seen that the value of the control input $v_{i2}$ of agent $i$ monotonically varies between its value before switching and that of its leader. Since leader's linear speed is bounded as per \eqref{input bound}, $u_i$ would also satisfy the constraint. For $\omega_{i2}$, the control gain $k_{\omega2}$ can be chosen as
\begin{align*}
    k_{\omega2} \leq \frac{1}{2\pi}(\omega_{max} - \omega_{lead(i)}),
\end{align*}
Above analysis renders $\omega_{lead(i)} \leq \omega_{max}-\Delta$ since leader is in one of the above modes. Hence, we have that 
\begin{align}
    k_{\omega 2} = \frac{\Delta}{2\pi}.
\end{align}
With this value of control gain, the angular speed of agent $i$ in this mode satisfies $|\omega_{i2}|\leq \omega_{max}$. This proves that $v_{ij}(t)$ and $\omega_{ij}(t)$ satisfies the control input bounds for all time $t\geq 0$ in each mode $q_{ij}$.
\end{proof}

\section{Proof of Lemma \ref{Zeno S}}\label{app: zeno}
\begin{proof}
We use the following condition to show that there is no Zeno behavior on the switching surface:
\begin{align*}
    \dot S^{ij-}\dot S^{ij+} > 0,
\end{align*}
where $\dot S^{ij} = \nabla S^{ij}\dot{\bm x}_i$ is the time derivative of $S^{ij}$ along the system trajectory and $-, +$ denotes the value of the derivative just before and just after the switch, i.e., on either side of the switching surface. This would imply that the vector field of the agent point in the same direction on either side of the surface (See Figure \ref{fig:switch surf}). 
\begin{figure}[!htbp]
    \centering
    \includegraphics[width=0.5\columnwidth,clip]{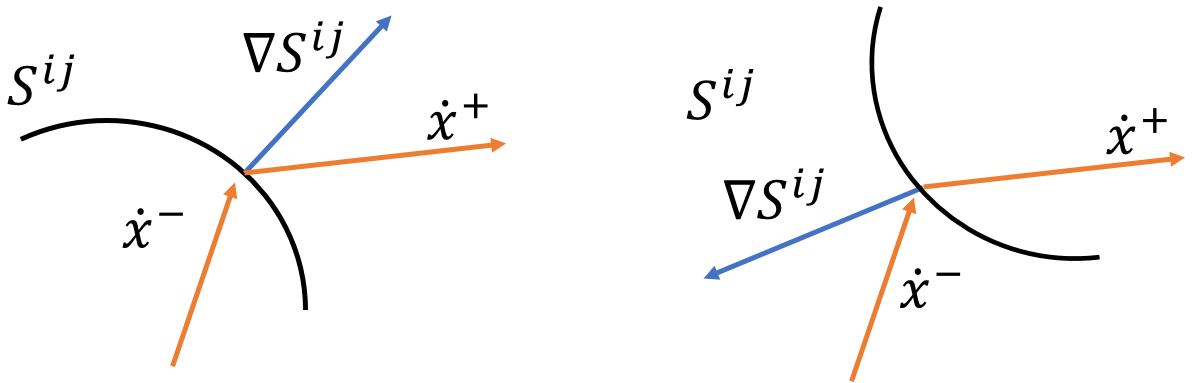}
    \caption{Geometric condition to avoid Zeno behavior.}\label{fig:switch surf}
\end{figure}
We analyze each switching surface as following: 

$S^{41}(i)$: Time derivative of $S^{41}$ reads:
\begin{align*}
    \dot{S}^{41}(i)^- = 2(\bm r_i- \bm r_j)^T(\dot{\bm r}_{i4} - \dot{\bm r}_{j4}) \; ; \quad  \dot{S}^{41}(i)^+ = 2(\bm r_i- \bm r_j)^T(\dot{\bm r}_{i1} - \dot{\bm r}_{j1}) 
\end{align*}
where $\dot{\bm r}_{i4} = \begin{bmatrix}v_{i4}\cos\theta_i &  v_{i4}\sin\theta_i\end{bmatrix}^T$ and  $\dot{\bm r}_{i1} = \begin{bmatrix}v_{i1}\cos\theta_j& v_{i1}\sin\theta_j\end{bmatrix}^T$. Since agents are moving towards each other just before they reach the switching surface, we have that $S^{41}(i)^- = 2(v_{i4}\bm r_{ij}^T\bm \eta_i - v_{j4}\bm r_{ij}^T\bm \eta_j)<0$ where $\bm \eta_i = \begin{bmatrix}\cos\theta_i & \sin\theta_i\end{bmatrix}^T$. The derivative $\dot{S}^{41}(i)^+ <0$ as the distance between the agents keep decreasing until they reach the circular orbit. This gives us $\dot S^{41-}\dot S^{41+} > 0$ which means there is no chattering on this surface. Also, the guard conditions $G(q_{i1},q_{i4})$ and  $G(q_{i4},q_{i1})$ are satisfied together only if $(\bm r_i-\bm r_j)^T\dot{\bm r}_i = 0$. In this case, agents $i$ and $j$ switch back to modes $q_{i4}$ and $q_{j4}$ respectively, just after switching to mode $q_{i1}$ and stay there.

$S^{43}(i)$: Its time derivative reads $S^{43}(i)^- = 2(\bm r_i- \bm r_j)^T(\dot{\bm r}_i^- - \dot{\bm r}_j^-)$ which is also negative since the agents are moving towards each other before reaching the switching surface. Furthermore, once on the other side of switching surface, i.e., once in mode $q_{i3}$ and $q_{j3}$, respectively, both the agents hold their direction of motion, which makes $2(\bm r_i- \bm r_j)^T(\dot{\bm r}_i^- - \dot{\bm r}_j^-) = 2(\bm r_i- \bm r_j)^T(\dot{\bm r}_i^+ - \dot{\bm r}_j^+)<0$.
Again, we get that $\dot S^{43-}\dot S^{43+} > 0$.
Therefore, there is no Zeno behavior on this surface. 

$S^{14}(i)$: The switching surface is given by $S^{14}(i): |\theta_i-\angle(\bm r_{gi} - \bm r_i)|-\delta = 0$ and its time derivative in the respective modes are given as:
\begin{align*}
    \dot{S}^{14}(i)^-  = \pm(\omega_{i1} - \frac{d}{dt}\angle(\bm r_{gi} - \bm r_i)) \; ; \quad \dot{S}^{14}(i)^+ = \pm(\omega_{i4} - \frac{d}{dt}\angle(\bm r_{gi} - \bm r_i)).
\end{align*}
Here $\pm$ depends upon the sign of $\theta_i-\angle(\bm r_{gi}-\bm r_i)$. Agent switch to this mode when $|\theta_i-\angle(\bm r_{gi} - \bm r_i)|-\delta \leq 0$. This means that at switching surface, the derivative $(\omega_{i1} - \frac{d}{dt}\angle(\bm r_{gi} - \bm r_i))$ must be negative since in mode $q_{i1}$, agent moves in a circular orbit, which means the difference in its orientation and the vector joining its current location and goal location ($|\theta_i-\angle(\bm r_{gi} - \bm r_i)|$) decreases, hence $\dot{S}^{14}(i)^- <0$. Now, on the other side of the switching surface, i.e., in mode $q_{i4}$, agent aligns its orientation along that of the vector $(\bm r_{gi}-\bm r_i)$ as per \eqref{global-attract}, which means the derivative of ($|\theta_i-\angle(\bm r_{gi} - \bm r_i)| = \dot{S}^{14}(i)^+ <0$). This implies  $\dot S^{14-}\dot S^{14+} > 0$ and hence, there is no Zeno behavior.

$S^{24}(i)$: The switching surface is given by $S^{24}(i) : (\bm r_{gi} - \bm r_i)^T(\bm r_{lead(i)} - \bm r_i) = 0 $. Hence
\begin{align*}
    \dot S^{24}(i) & = -\dot{\bm r}_{i2}^T(\bm r_j - \bm r_i) +(\bm r_{gi} - \bm r_i)^T(\dot{\bm r}_j - \dot{\bm r}_{i2})    
    % & = -\dot{\bm r}_{i2}^T(\bm r_j + \bm r_{gi}- 2\bm r_i) +(\bm r_{gi} - \bm r_i)^T\dot{\bm r}_j
\end{align*}
The switching surface represents the angle between the vectors $\bm r_{gi}-\bm r_i$ and $\bm r_{lead(i)}-\bm r_i$. While in mode $q_{i2}$, agent switches to mode $q_{i4}$ only if the goal location is on the free side. Hence, in mode $q_{i2}$, the angle between these 2 vectors would be increasing. Once the agent switches, it starts moving towards its goal location. This would further increase the angle between the 2 vectors. Hence, the time derivative $\dot S^{24}(i)$ is positive on either side of the surface. Also, due to the guard condition, agent makes a switch only when its goal location is on the free side. Therefore, agent $i$ moves away from the leader $lead(i)$ and it would not come in conflict with the formation again and hence, there would be no Zeno behavior.

$S^{32}(i)$: The switching surface is given by  $S^{32}(i) : \|\bm r_i - \bm r_k\|- R_c^2 =  0 $. Before switching to mode $q_2$, agents $i$ and $k$ are moving towards each other, i.e., $\dot S^{32}(i)^- =  2(\bm r_i- \bm r_k)^T(\dot{\bm r}_{i3} - \dot{\bm r}_{k}) < 0$. Once it switches, it changes its speed and orientation to match that of its leader. Until then, the distance between the 2 agents keeps decreasing (see Lemma \ref{FL safe}). Hence, the time derivative on the other side $\dot S^{32}(i)^+ < 0$. 
Hence, agent stays on other side of switching surface.

$S^{34}(i)$: The switching surface is governed by $S^{34}(i): v_{i3}(\bm r_i-\bm r_j)^T\bm \eta_i - v_{j3}(\bm r_i-\bm r_j)^T\bm \eta_j = 0$. This is equal to the rate change of inter-agent distance between agent $i$ and $j$. Since the switching occurs when $\dot d_{ij}$ becomes positive, the derivatives on either side of the surface are, i.e., $\dot{S}^{34}(i)^- = \ddot d_{ij}^-$ and $\dot{S}^{34}(i)^+ =\ddot d_{ij}^+ $ are positive which implies no Zeno behavior on this switching surface.
\end{proof}

\end{document}